\documentclass[11pt]{article}

\usepackage[margin=1in]{geometry}
\usepackage[T1]{fontenc}
\usepackage[utf8]{inputenc}
\usepackage{lmodern}
\usepackage{amsmath,amssymb,amsfonts}
\usepackage{booktabs}
\usepackage{caption}
\usepackage{subcaption}
\usepackage{graphicx}
\usepackage{hyperref}
\usepackage{siunitx}
\usepackage{multirow}
\usepackage{xcolor}
\usepackage{array}
\usepackage{mathrsfs}
\usepackage{makecell}
\usepackage{physics}
\usepackage{authblk}

\newcommand{\teal}[1]{{\color{teal}#1}}

\hypersetup{
    colorlinks=true,
    linkcolor=blue,
    citecolor=blue,
    urlcolor=blue,
}

\graphicspath{{figures/}}

\title{Model-Based Fast Calibration for a Nitrogen Vacancy Center QPU}
\author[1]{Tobias F. Sjolander\textsuperscript{$\dagger$}}
\author[1]{Marco Rossignolo\textsuperscript{$\dagger$}}
\author[1]{Lorena C. Bianchet\textsuperscript{$\dagger$}}
\author[1]{Swathi Hegde}
\author[1]{Anurag Saha Roy}
\author[2,3]{Thomas Lippert}
\author[2]{Manpreet Singh Jattana}

\affil[1]{Qruise GmbH, 66113 Saarbrücken, Germany}

\affil[2]{Modular Supercomputing and Quantum Computing,
Goethe University Frankfurt,
Kettenhofweg 139,
60325 Frankfurt am Main, Germany}

\affil[3]{Jülich Supercomputing Centre,
Institute for Advanced Simulation,
Forschungszentrum Jülich,
Wilhelm-Johnen-Straße,
52428 Jülich, Germany}

\date{\today}

\begin{document}

\maketitle

\begingroup
\renewcommand{\thefootnote}{\fnsymbol{footnote}}
\footnotetext[2]{These authors contributed equally to this work.}
\endgroup

\begin{abstract}
We present a simulation-informed calibration workflow for NV quantum processing units that combines a digital twin, optimal control pulses, and a Fisher-information-guided experiment design into an iterative parameter estimation feedback loop. 
We find that using this framework, we obtain the same calibration precision as standard techniques with up to 4x reduction in measurement time.
\end{abstract}

\section{Introduction}
Calibration is a key component of practical quantum technology workflows, both for quantum sensing, and quantum computation. Signal generator voltages, piezo settings, and amplifier currents must be reliably mapped to internal Hamiltonian parameters in order to calculate drive envelopes, flux bias settings and other relevant quantities. In quantum computing, such settings are crucially important to gate fidelity~\cite{krantz2019quantum,bruzewicz2019trapped,evered2023high,bluvstein2022quantum}. Whereas in quantum sensing, obtaining the Hamiltonian parameters of some model quantum system for a given environmental stimulus is the whole point - these parameters correspond to the measurement results.

Commonly implemented procedures rely on repeated sweeps over control parameters such as drive frequency, pulse duration, or inter-pulse timing, followed by analytical fitting routines using the predicted response to the implemented sweep. This process is robust and simple, but it uses laboratory time inefficiently. In particular, for a series of independent measurements, using the same control parameters, the signal-to-noise ratio (SNR) of the measurement,and therefore the accuracy of the analytical fit, improves as $\mathrm{SNR}\propto\sqrt{N}$, where $N$ is the number of sweep repetitions. Further, sequential calibrations tend to ignore shared physics. For example, a typical workflow would involve calibrating the resonance frequency of a qubit by sweeping the frequency of some sort of drive across the resonance, followed by a Rabi type experiment where the drive is parked at the determined resonance value and the amplitude or duration is swept in order to calibrate the coupling strength between the qubit and the drive. 
In this case, the potential correlation between the resonance linewidth and the coupling strength is often discarded as a source of information since the dependency may be difficult to calculate exactly. Finally, there is also the fact that not all measurement conditions are equally informative. 
For example, in many experiments, measurements taken near the apparent center of a resonance or away from the steep parts of a signal carry little information about the latent parameters that must be estimated. A calibration loop that continues to probe such regions will converge slowly.

Bayesian calibration and system-identification methods provide a principled route to efficient quantum-device characterization by maintaining a probability distribution over unknown parameters and updating it as measurements are acquired. The posterior can be represented, for example, on a parameter grid, by particles, or through a variational approximation; the updated uncertainty can then guide the selection of subsequent measurements. Bayesian approaches have been developed for dissipative two-level-system identification and optimized adaptive system identification~\cite{schirmer2015UbiquitousProblemLearning,stace2022OptimisedBayesianSystem}, and have been demonstrated for trapped-ion entangling-gate calibration and multidimensional NV-center model learning~\cite{gersterExperimentalBayesianCalibration2022,belliardo2026MultidimensionalQuantumEstimation}. Related active-learning schemes use surrogate models to reduce the cost of evaluating expensive calibration objectives~\cite{walton2026ActiveLearningCalibrating}. In practice, the computational cost of Bayesian calibration can grow rapidly with the number of inferred parameters, the resolution required to represent the posterior, and the cost of repeated forward-model evaluations.\\

In this work, we introduce a closed loop model-based parameter estimation workflow intended to resolve several of the challenges outlined above, together with an implementation of this idea. The workflow combines three components: a physics-based digital twin, a model-learning (ML) feedback loop that refits model parameters from each new batch of measurements, and a Fisher-information (FI) guided experiment design that chooses the next measurement settings based on the current model state at each iteration of the loop. The digital twin enables detailed simulations of the system response to arbitrary controls, ensuring that cross correlations between parameters and shared physics are captured natively. Further, the twin allows us to calculate gradients of the observables with respect to the measurement settings, which gives us access to the Fisher information. Meanwhile, the loop based design allows the estimate for the Fisher-information matrix to improve throughout the experiment as long as the estimates for the model parameters converge, meaning that the measurement settings get progressively more informative as the calibration proceeds. This implies that such a workflow enables faster than $\sqrt{N}$ scaling of SNR with experiment time, unless the control points are already maximally informative, which would imply that the system is already characterized and the model parameters are already known.

The paper includes both a detailed description of the algorithm as well as numerical and experimental demonstrations. For the experiments we use an NV centre system, although the method can be applied to any system for which accurate modeling is possible. 

This work is structured as follows:
In the Section \ref{sec:calibration_workflow} we explain the details of the proposed calibration feedback loop and its different constituent parts. In Section \ref{sec:methods} we describe the details of the experimental setup and the theory behind the modeling.
In the Section \ref{sec:results}, we present the results from our implementation of this framework on an NV QPU and show how it enables efficient parameter estimation in practice. We prove the goodness of the parameters learnt via our calibration workflow against standard calibration technique by benchmarking with quantum process tomography (QPT) and randomized benchmarking (RB). Finally we discuss the implications of this work and the possible future directions \ref{sec:conclusions}.
\section{Calibration Workflow}
\label{sec:calibration_workflow}

\begin{figure}[h]
    \centering
    \includegraphics[width=0.8\linewidth]{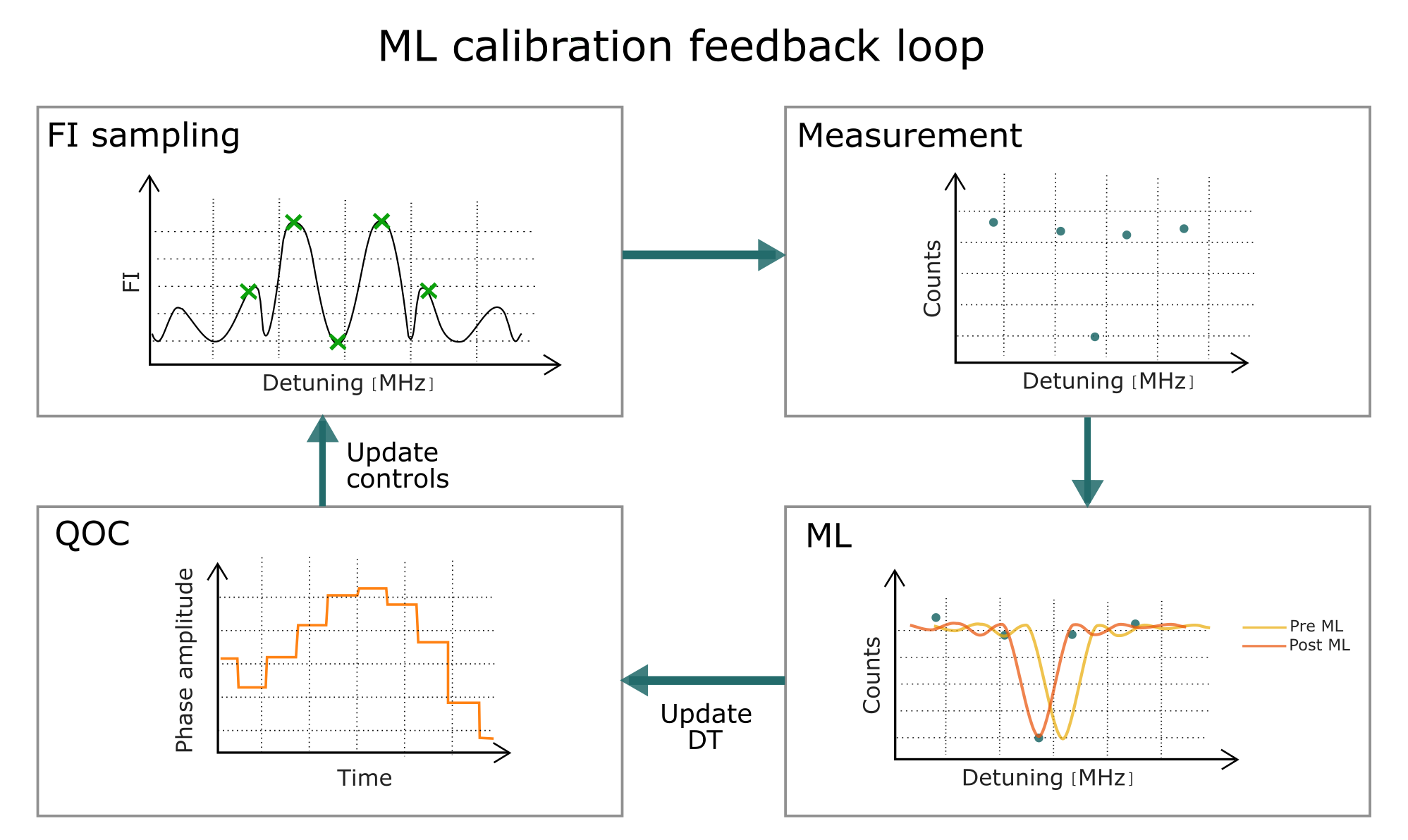}
    \caption{\textbf{Model-based calibration feedback loop.} The four connected panels summarize the iterative workflow used throughout this work. In the QOC panel, the current digital-twin parameters are used to optimize the control pulse so that the implemented sequence produce the best outcome from the device. In the FI sampling panel, that updated pulse is held fixed while candidate experimental settings are scored according to how informative they are expected to be for the unknown model parameters, and the most informative settings are selected for the next experiment. In the measurement panel, the chosen control sequence is executed on the NV QPU and the resulting data are acquired. In the ML panel, the digital-twin parameters are updated by fitting the model to the accumulated measurement record. The updated parameter estimate is then fed back to the QOC stage, closing the loop. The same model therefore serves both retrospective inference, by learning the latent physical parameters from data, and prospective experiment design, by updating the control pulse and selecting the next measurement locations.}
    \label{fig:01_methods}
\end{figure}

Our proposed calibration workflow is summarized schematically in Figure~\ref{fig:01_methods}. For a given experiment design, consisting of some sequence of gates, we start by building a digital twin of the system to be calibrated, and make an initial guess for the model parameters to be learned. Next, we use quantum optimal control (QOC) theory to calculate the drive envelopes that implement the desired gates given the current knowledge of the system. Next, we use a quantitative scoring method based on the Fisher information~\cite{fisher1925theory} to calculate the maximally informative measurement conditions. Note that both the pulse shape calculation and the measurement scoring are sensitively dependent on the model parameters.
Next, we perform the calibration experiment using the optimized pulses, at the determined settings, and obtain a system response. We use gradient descent learning to find the best-fit model parameters that minimizes the difference between the measured system response and the same measurement calculated using the digital twin. The best-fit model parameters are used to calculate a new set of control pulses and measurement conditions and the loop is closed.

\subsection{Digital Twin Setup}
\label{sec:DT_setup}
Our proposed calibration workflow relies on a sufficient microscopic model of the system. We call this model a digital twin, $\mathcal{DT}$. There is an obvious tradeoff between simplicity and computational speed on one hand, and model capacity and accuracy on the other.
The digital twin is parametrized by $\boldsymbol{\Theta}$, with some subset $\boldsymbol{\theta}$ being the $m$ parameters that are considered model parameters and are the target for learning or calibration.

The calibration workflow further requires us to define an 'experiment', $\mathcal{E}$, or sequence of control pulses to use for the calibration loop and program the corresponding physics into the digital twin. The goal when designing $\mathcal{E}$ is to generate an experimental signal that can be adequately modeled given the size of the model we have chosen $\boldsymbol{\Theta}$, and further, is strongly variable with respect to $\boldsymbol{\theta}$.\\
$\mathcal{E}$ is characterized by a set of settings $\boldsymbol{u}$, these can be, for example, wait times between pulses, or voltage settings on signal generators. The important distinction between $\boldsymbol{\theta}$ and $\boldsymbol{u}$ is that we have presumed full experimental control over $\boldsymbol{u}$. It is worth making explicit that a particular term in the model Hamiltonian, say the magnetic bias field, can be either a parameter, or a setting, depending on the design of $\mathcal{DT}$ and $\mathcal{E}$.\\
The measured signal resulting from an experiment with particular settings is modeled as $x_{\mathcal{E}} = \mathcal{DT}_{\mathcal{E},\boldsymbol{\Theta}}(\boldsymbol{\theta},\boldsymbol{u}, \boldsymbol{c})$, where $\boldsymbol{c}$ denotes variables associated with pulse shaping.

\subsection{Pulse shaping - QOC}
\label{subsec:QOC}
For a general quantum system, there is no analytical solution for an arbitrary gate. This means that obtaining a process with high fidelity can be challenging and the learning landscape can manifest degenerative solutions. For this reason, the goal of the first step is to adapt the drive envelopes to the current knowledge of the system through optimal control theory~\cite{ansel2024OptimalControlStrategies}.
To carefully distinguish pulse synthesis from measurement placement, since both $\boldsymbol{u}$ and $\boldsymbol{c}$ depend on $\boldsymbol{\theta}$, it is useful to separate the sequence $\mathcal{E}$, into blocks composed of a control stage and a free evolution stage,
\begin{equation}
U(\boldsymbol{\theta},\boldsymbol{u},\boldsymbol{c})=U_{\mathrm{control}}(\boldsymbol{\theta},\boldsymbol{u},\boldsymbol{c})U_{\mathrm{free}}(\boldsymbol{\theta},\boldsymbol{u}).
\end{equation}
here $U_{\mathrm{control}}(\boldsymbol{\theta},\boldsymbol{u},\boldsymbol{c})$ is the propagator for the drive pulse while $U_{\mathrm{free}}(\boldsymbol{\theta},\boldsymbol{u})$ is the propagator for free evolution. This separation is useful because the control and the experimental settings play complementary roles. The control propagator will be used during optimization, whereas the full propagator will be used in the next step of the workflow to find the ideal measurement settings. 
The control stage is obtained by optimizing a fidelity objective under the current model knowledge. If $U_{\mathrm{target}}$ denotes the target gate, a standard control objective is the infidelity
\begin{equation}
\mathcal{L}_{\mathrm{QOC}}(\boldsymbol{\theta},\boldsymbol{\tilde{u}},\boldsymbol{c})=1-F\!\left(U_{\mathrm{control}}(\boldsymbol{\theta},\boldsymbol{\tilde{u}},\boldsymbol{c}),U_{\mathrm{target}}\right),
\label{eq:fidelity-loss}
\end{equation}
where $F$ is the chosen fidelity metric. In the present workflow, the control pulse is optimized at a fixed nominal operating point $\boldsymbol{\tilde{u}} \in \mathcal{U}$ the set of achievable measurement settings. We adopt this convention because we determine the measurement settings in the next step by maximizing the overall information of the sequence. This distinction keeps the model parameters and control variables separated from the measurement settings.\\
The pulse-design step takes the form
\begin{equation}
\boldsymbol{c}^{\star}(\boldsymbol{\theta},\boldsymbol{\tilde{u}} ) \in \arg\min_{\boldsymbol{c}\in\mathcal{C}} \mathcal{L}_{\mathrm{QOC}}(\boldsymbol{\theta},\boldsymbol{\tilde{u}} ,\boldsymbol{c}).
\label{eq:control-learning}
\end{equation}
This produces the optimized control propagator $U_{\mathrm{control}}(\boldsymbol{\theta},\boldsymbol{u},\boldsymbol{c}^{\star}(\boldsymbol{\theta},\boldsymbol{\tilde{u}}))$ that is then embedded into the full measurement sequence.

\subsection{Measurement selection - Fisher information}
\label{sec:fi-selection}
In the second step of the calibration workflow, the digital twin selects experimental settings $\boldsymbol{u}$. Remember that $\boldsymbol{u}$ denotes a particular choice about how to carry out the experiment, for example a given microwave detuning together with a particular inter-pulse wait time. What options are available for $\boldsymbol{u}$ depends on the hardware and the design of the specific experiment $\mathcal{E}$. A key part of our proposed workflow is that these settings are chosen to be maximally informative given current model knowledge with the optimized pulse $\boldsymbol{c}^{\star}(\boldsymbol{\theta},\boldsymbol{\tilde{u}})$ held fixed.
The design score for 'informative' used in this work is the classical Fisher information. For model parameters $\boldsymbol{\theta}$, experimental settings $\boldsymbol{u}$, and optimized pulse shaping variables $\boldsymbol{c}^{\star}$, let $p(\boldsymbol{u}, \boldsymbol{\theta},\boldsymbol{c}^{\star})$ denote the predicted population in the measured basis. The corresponding coefficients of the FI matrix $\boldsymbol{\mathcal{I}}$ are
\begin{equation}
\mathcal{I}_{ab}(\boldsymbol{u},\boldsymbol{\theta},\boldsymbol{c}^{\star}) = V_{p}(p) \,\frac{\partial p}{\partial \theta_{a}} \frac{\partial p}{\partial \theta_{b}}
\label{eq:scalar-fi}
\end{equation}
where $(a,\, b)$ are the indices of the model parameters, $p \equiv p(\boldsymbol{u},\boldsymbol{\theta},\boldsymbol{c}^{\star})$ to simplify the notation, and $V_{p}(p)$ is a function depending on the noise model chosen to describe the system behavior (see App. \ref{app:fisher-information-model}).

Equation~\eqref{eq:scalar-fi} explains why intuitive operating points are often poor calibration points. As an example, a measurement taken at the peak of a resonance may be experimentally natural, but since the derivative with respect to the resonance frequency goes to zero there, that measurement contributes little information about parameters that determine the resonance frequency, such as the magnetic field in the case of NVs. This fact is commonly exploited in magnetometry~\cite{taylor2008SensitivityOptimizationNV,degen2017QuantumSensing,sone2017HamiltonianIdentifiabilityAssisted}: the ideal measurement points for determining the position of a resonance line are on the wings of the Lorentzian lineshape, where the derivative with respect to magnetic field is maximum, rather than at resonance.
Eq.~\eqref{eq:scalar-fi} is the multidimensional generalization of this point to arbitrary parameters, while also accounting for the variance arising from read out noise.\\
In practice, in our implementation, the digital twin selects a set of several distinct measurement settings $\lbrace \boldsymbol{u} \rbrace$ using a design score based on a scalar summary of the FI matrix for each iteration of the calibration loop. To be more specific, for this implementation we chose a design score that maximises the determinant of the sum of the FI matrices for the set of chosen measurement settings. For each measurement point $i$ and total budget $M$ of measurement points, the design score $S$ is defined as

\begin{equation}
    S = \operatorname{det}\left(\sum_{i}^{M}\mathcal{I}(\boldsymbol{u}_i,\boldsymbol{\theta},\boldsymbol{c}^{\star})\right)
\end{equation}
and an optimization over the family $\lbrace \boldsymbol{u} \rbrace = \left( \boldsymbol{u}_1, \dots, \boldsymbol{u}_M \right)$ as,
\begin{equation}
\lbrace\boldsymbol{u}^{\star}\rbrace \in \arg\max_{\boldsymbol{u}\in\mathcal{U}} S(\lbrace\boldsymbol{u}\rbrace;\boldsymbol{\theta},\boldsymbol{c^{\star}}(\boldsymbol{\theta},\boldsymbol{\tilde{u}}))
\label{eq:fisher-optimization}
\end{equation}

Each individual measurement setting $\boldsymbol{u}$ is selected to maximize the joint sensitivity of the measurement to the parameters $\boldsymbol{\theta}$ to be learned, given all the previously selected settings for that iteration.
For each new setting to be chosen, we evaluate Eq.~\eqref{eq:scalar-fi} over a grid of allowed settings, subject to the relevant experimental constraints, and the next $\boldsymbol{u}$ is chosen such that $S$ is maximal, as described in App.~\ref{app:doptimal}. The number of distinct measurement settings per iteration, as well as the density and size of the grid over which they are evaluated, become hyperparameters of the calibration loop. Note that this selection process is carried out using a control pulse previously optimized in the the first step of the workflow, with fixed nominal settings $\boldsymbol{\tilde{u}}$, as described in Sec.~\ref{subsec:QOC}.

\subsection{Measurement and model learning}
Next we obtain the physical measurements $\lbrace y \rbrace$ corresponding to $\mathcal{E}$, $\lbrace \boldsymbol{u} \rbrace$, and $\boldsymbol{c}^\star$.

The model parameters are updated based on the accumulated measurement history. At a particular iteration, $k$, we have a batch, $B_k$, that collects the measurement settings, the controls, and the measurement results:
\begin{equation}
B_k = (\lbrace \boldsymbol{u^{\star}} \rbrace_k, \boldsymbol{c^{\star}}_k, \lbrace y\rbrace_k)
\end{equation}  
where $\lbrace y \rbrace _k$ is the set of scalar measurement data resulting from the k-th set of measurement settings $\lbrace \boldsymbol{u^{\star}} \rbrace_k$.
For each measured $y$, the digital twin predicts the equivalent simulated observable $x$. Given all $x$ and $y$ for each batch the parameter update uses all the available batches $\mathcal{B}_k = \lbrace B_1, \dots,  B_k\rbrace$ and solve the optimization problem:
\begin{equation}
\boldsymbol{\theta}_{k+1} \in \arg\min_{\boldsymbol{\theta}}
\mathcal{L}_{\mathrm{ML}}(\boldsymbol{\theta};\mathcal{B}_k),
\label{eq:model-learning}
\end{equation}
where $\mathcal{L}_{\mathrm{ML}}$ is an objective function that quantifies the discrepancy between measured and simulated responses for all the measurements done in the calibration loop.

In our implementation, the objective is a least-squares loss,
\begin{equation}
\mathcal{L}_{\mathrm{ML}}(\boldsymbol{\theta};\mathcal{B}_k)
= \sum_{i \in \mathcal{B}_k} w_i\,\bigl(y_i - x_i(\boldsymbol{\theta})\bigr)^2,
\label{eq:learning-loss}
\end{equation}
where $y_i$ is the $i$-th measured observable, $x_i(\boldsymbol{\theta})$ is the corresponding digital-twin prediction at parameters $\boldsymbol{\theta}$, and the weights $w_i$ can encode shot counts or batch importance.
The objective function can vary across applications, but the loop-level logic is unchanged: refit the physical model to the current data, update the control pulse under that model, and then use the resulting sequence for experiment design.
The resulting closed loop alternates between updating the controls, choice of measurement settings (experiment design), measurement, and learning. After solving Eq.~\eqref{eq:model-learning}, the control pulse is first updated from the new parameter estimate by optimizing the fidelity objective in Eq.~\eqref{eq:fidelity-loss} at the nominal operating point $\boldsymbol{\tilde{u}}$, yielding the pulse parameters $c^{\star}(\boldsymbol{\theta}_{k+1},\boldsymbol{\tilde{u}})$. The next experiment is then chosen accordingly to Eq.~\eqref{eq:fisher-optimization}.
The same digital twin is therefore used both retrospectively to refit latent
parameters and prospectively to update the control pulse and the next experiment design under the current model.\\

\subsection{Parameter error estimation}
\label{sec:parameter-error-estimation}
After the model-learning loop has converged to an estimate $\boldsymbol{\hat{\theta}}$,
it is useful to attach standard deviation errors to each parameter to quantify the
remaining estimation uncertainty.
We adopt the Gauss--Newton covariance approach, which provides a computationally
efficient post-fit diagnostic that requires only one automatic-differentiation pass
through the already-converged twin.

Let $r_i(\boldsymbol{\theta}) = x_i(\boldsymbol{\theta}) - y_i$
denote the scalar residual at the $i$-th data point.
Stacking all $n$ residuals accumulated across all measurement settings gives the
residual vector
\begin{equation}
  \boldsymbol{r}(\boldsymbol{\theta})
  = \left(r_1(\boldsymbol{\theta}),\,\ldots,\,r_n(\boldsymbol{\theta})\right)
  \label{eq:residual-fn}
\end{equation}
The Jacobian of the residual vector with respect to the model parameters,
evaluated at $\boldsymbol{\hat{\theta}}$, is
\begin{equation}
  J_{ia}
  = \left.\frac{\partial r_i}{\partial \theta_a}\right|_{\hat{\boldsymbol{\theta}}} \in \mathbb{R}^{n\times m}
  \label{eq:jacobian}
\end{equation}
computed by automatic differentiation through the differentiable twin.
The unbiased residual variance is estimated as
\begin{equation}
  \sigma^2 = \frac{\mathrm{RSS}}{n - m},
  \qquad
  \mathrm{RSS} = \lVert \boldsymbol{r}(\hat{\boldsymbol{\theta}})\rVert_2^2,
  \label{eq:sigma2}
\end{equation}
where $n-m$ is the number of degrees of freedom.
The Gauss--Newton approximation to the parameter covariance~\cite{kay1993fundamentals} is then
\begin{equation}
  \operatorname{Cov}(\hat{\boldsymbol{\theta}})
  = \sigma^2\,\bigl(J^\top J\bigr)^{-1} \in \mathbb{R}^{m\times m},
  \label{eq:gauss-newton-cov}
\end{equation}
and the standard deviation error on parameter $a$ is
$\hat{\sigma}_a = \sqrt{[\operatorname{Cov}(\hat{\boldsymbol{\theta}})]_{aa}}$.

The matrix $J^\top J$ is the empirical Fisher information matrix accumulated over all experimental data: its $(a,b)$ entry is the
inner product of the $a$-th and $b$-th columns of $J$, measuring how
consistently the two parameters are distinguished across the full measurement
record.
Well-conditioned Fisher-information-guided measurement settings (Section~\ref{sec:fi-selection})
directly ensures that $J^\top J$ is non-degenerate, making the covariance
estimate reliable.
Conversely, a uniform or poorly designed grid may leave $J^\top J$
nearly singular; in that case the Moore--Penrose pseudoinverse is used in place
of the ordinary inverse, so that poorly constrained directions contribute zero
rather than infinite variance to the diagonal.

The estimate in Eq.~\eqref{eq:gauss-newton-cov} rests on three assumptions.
First, the noise is homoscedastic: a single variance $\sigma^2$ applies to all
residuals.
Second, the Gauss--Newton approximation is valid: the calibration has converged
to a low-residual minimum so that the second-order correction to the full
nonlinear covariance is negligible.
Third, the asymptotic regime applies: $n \gg p$, so that $\hat{\sigma}^2$ is a
reliable estimate of the true noise variance.
Under these conditions the standard deviation errors shrink as $\hat{\sigma}_a \propto
1/\sqrt{n}$, consistent with the Cram\'{e}r--Rao bound.

\section{Methods}
\label{sec:methods} 
As a proof-of-principle demonstration, we use the proposed method to calibrate 1Q gates of a QPU based on NV centers in diamond. NV centers are of interest in this context due to their popularity as general quantum technology platform~\cite{doherty2013nitrogen,degen2017QuantumSensing}. 

\subsection{NV-center system and digital twin model}
In this work, we construct a digital twin of a room-temperature NV center QPU based on quantum simulation of spin degrees of freedom. The model uses a JAX based framework to enable auto-differentiation with respect to model parameters, control variables, and measurement settings. We model the readout by assigning a fixed count rate to each spin state.

Following the standard treatment ~\cite{doherty2013nitrogen} the Hamiltonian for an NV center coupled to a $^{14}$N nucleus, with coupled $^{13}$C spins is:
\begin{align}
H &= \underbrace{D\!\left(S_z^2 - \tfrac{S(S+1)}{3}\right) + E\!\left(S_x^2-S_y^2\right)}_{H_{\rm ZFS}}
    - \underbrace{\gamma_e\,\boldsymbol{B}_0\cdot\boldsymbol{S}}_{H_{\rm Z}^{(e)}} \nonumber\\
  &\quad + \underbrace{A_\parallel\,S_z I_z^{(N)} + A_\perp\!\left(S_x I_x^{(N)}+S_y I_y^{(N)}\right)}_{H_{\rm hf}^{(N)}}
    + \underbrace{P\!\left[\!\left(I_z^{(N)}\right)^{\!2} - \tfrac{I^{(N)}(I^{(N)}+1)}{3}\right]}_{H_{\rm quad}^{(N)}}
    - \gamma_N\,\boldsymbol{B}_0\cdot\boldsymbol{I}^{(N)} \nonumber\\
  &\quad + \sum_k\!\left[\boldsymbol{S} \cdot \boldsymbol{A}^{(C_k)} \cdot \boldsymbol{I}^{(C_k)}
    - \gamma_C\,\boldsymbol{B}_0\cdot\boldsymbol{I}^{(C_k)}\right],
\label{eq:H_NV}
\end{align}
where $\boldsymbol{S}=(S_x,S_y,S_z)$ is the $S=1$ NV electron spin operator,
$\boldsymbol{I}^{(N)}$ is the $^{14}$N nuclear spin ($I=1$),
$\boldsymbol{I}^{(C_k)}$ is the $k$-th $^{13}$C nuclear spin ($I=1/2$),
$D\approx 2\pi\times 2.87\,\mathrm{GHz}$ is the zero-field splitting,
$E$ is the strain-induced transverse splitting (often negligible),
$\gamma_e/2\pi\approx 28\,\mathrm{GHz/T}$ and $\gamma_{N,C}$ are the
electron and nuclear gyromagnetic ratios,
$A_\parallel,A_\perp$ are the axial and transverse $^{14}$N hyperfine
couplings, $P\approx -2\pi\times 4.95\,\mathrm{MHz}$ is the $^{14}$N
electric quadrupole parameter, $\boldsymbol{A}$ is the hyperfine tensor that defines the hyperfine couplings $A_{\parallel,k} = A_{zz,k}$, $A_{\perp,k} = \sqrt{A^{2}_{xz} + A^{2}_{yz}}$ of the
$k$-th $^{13}$C spin.
The microwave drive adds the time-dependent term
\begin{equation}
H_{\rm drive}(t) = \sqrt{2}\,\Omega(t)\cos\!\bigl[\omega_d\,t + \phi(t)\bigr]\,S_x,
\label{eq:H_drive}
\end{equation}
where $\Omega(t)$ and $\phi(t)$ are the Rabi-frequency and phase envelopes and $\omega_d$ is the microwave drive frequency.

\subsection{Experimental setup}
\label{subsec:experimental}
The QPU used for the measurements was a XQ1 Quantum Processor System by XeedQ. The setup is divided in two: (a) an optical table enclosed in a box of dimensions $45$ cm $\times 50$ cm $\times 15$ cm (L$\times$ W$\times$ H) located in an isolated room and, (b) control electronics placed in an adjacent room connected via cables to the optics table. The optical table is not actively cooled and remains in equilibrium with the room temperature, which changes at a slow and continuous rate. The QPU consists of a single 14N NV-center weakly coupled to neighboring 13C spins. The QPU is operated at the NV excited state level anti-crossing (ESLAC) by means of a bias magnetic field of $0.05$ T applied along the defect axis. State preparation of the NV and 14N spins is done by applying green light, which at the ESLAC initializes the NV electron and the coupled 14N spin into the joint $\ket{m_S=0, m_I=1}$ spin state, where $m_S$($m_I$) denotes the electron(nuclear) spin projection along the defect axis \cite{HARRISON2004Lumi,Jacques2009PRL}

\begin{figure*}[t]
  \centering
  \includegraphics[width=\textwidth]{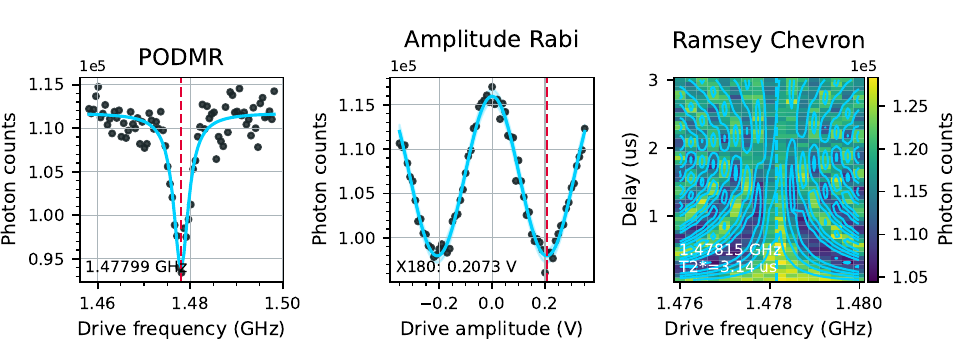}
  \caption{
    \textbf{Standard calibration} Example calibration flow for 1Q gates on the NV electron.
  }
  \label{fig:nv1-latest-calibrations}
\end{figure*}

During normal operation the QPU is calibrated daily using a linear flow of experiments. 1Q gates on the NV electron are calibrated by pulsed optically detected magnetic resonance (podmr) to determine an initial estimate of the qubit frequency, followed by a rabi experiment to determine the voltage to rotation angle conversion. Finally the qubit frequency is fine-tuned with a ramsey chevron experiment using the updated pulse parameters. In addition, the ramsey chevron experiment yields the dephasing time $T_2^\star$. We show an example of such a flow in Figure~\ref{fig:nv1-latest-calibrations}. During normal operation, gate parameters obtained from such calibration flows are stored in a Knowledge Base (KB) that serves as a source of truth for the state of the QPU at any given moment. 

\section{Results}
\label{sec:results}
In the following, we present proof-of-principle implementations of our proposed calibration workflow using the digital twin and QPU setup described in the previous section. We obtain various model parameters of interest and verify that using these we are able to accurately predict measurement results corresponding to different experiment designs. Further, we use the learned model parameters to calculate gates for the QPU and verify that the gate-fidelity is matched between the gates obtained from loop based calibration and gates obtained from the normal calibration flow based on analytic fitting routines. 

\subsection{Pulse Sequences}
\begin{figure}[h!]
    \centering
    \begin{minipage}{0.48\textwidth}
        \centering
        \includegraphics[width=\linewidth]{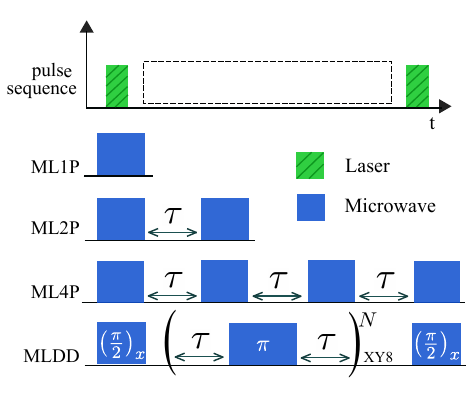}
    \end{minipage}
    \hfill
    \begin{minipage}{0.48\textwidth}
        \centering
        \begin{tabular}{lcc}
            \toprule
            Sequence & \makecell{Learned\\parameters} & \makecell{Measurement\\settings}\\
            \midrule
            ML1P & $B_0$ $\Omega$ & $\omega_d$ \\
            ML2P & $B_0$ $\Omega$ $\gamma$ $\tau_p$ & $\omega_d$ $\tau$ \\
            ML4P & $B_0$ $\Omega$ & $\omega_d$ $\tau$ \\
            MLDD & $A_{\mathrm{C13},\parallel}$ $A_{\mathrm{C13},\perp}$ & $\tau$ $N$\\ 
            \bottomrule
        \end{tabular}
    \end{minipage}
    \caption{\textbf{Experiments design overview.} Left: Scheme of the pulse sequence used in the experiment: ML1P, ML2P, ML4P and MLDD sequence. Right: table with description of the learned parameters and the measurement settings for each sequence. $B_0$ is the magnetic field aligned with the NV axis , $\Omega$ the Rabi frequency, $\gamma$ the dephasing rate, $\tau_p$ corresponds to an unknown delay time after each pulse, $A_{\mathrm{C13},\parallel}$ $A_{\mathrm{C13},\perp}$ are hyperfine couplings, $\omega_d$ is the drive frequency. }
    \label{fig:sequences}
\end{figure}

An important aspect of any calibration is that the controls or pulse sequence applied to the system should produce a response that varies strongly with the parameters to be calibrated. To demonstrate the generality of our method, we implement the calibration loop for several different pulse sequences of increasing complexity, including a single phase modulated microwave burst as well as a long periodic pulse train containing 100s of $\pi$-pulses figure~\ref{fig:sequences}. The 1-pulse sequence (ML1P) is used to learn the Rabi frequency and a resonance-related field parameter while optimizing probe placement over drive-frequency offsets. The 2-pulse sequence (ML2P) expands the control space from one dimension to two, making the experiment design problem richer as the twin identifies informative points in a two-dimensional control space. In this case also the amount of learned parameters is increased. The 4-pulse sequence (ML4P) extends the two-pulse one by adding two additional free-evolution times. The dynamical decoupling (MLDD) sequence is used to characterize weakly coupled nuclear spins \cite{Zhao2012Sensing}. In this section we describe the calibration loop and results with an example for the single-pulse sequence, the other experiments can be found in the Appendix \ref{Appendix:other_sequences}.

\subsection{ML1P sequence implementation and results}
The simplest implementation in this work uses a single fixed amplitude, phase-modulated microwave pulse. We take $\omega_q$, the qubit frequency, as well as the Rabi frequency $\Omega$, the rotation rate of the NV spin at the pulse amplitude, as the model parameters to be learned. The relevant measurement setting for this experiment is the MW drive frequency.
The target objective for the pulse shaping is population transfer of the NV spin from the $m_S=0$ state to the $m_S = -1$ state. The pulse was chosen to have a duration x1.8 longer than an on-resonance $\pi$-pulse of the same amplitude - in order to enable inversion solutions across the entire bandwidth of interest.
Following the process described in Sec. \ref{sec:calibration_workflow}, we re-calculate the phase envelopes $\phi(t)$ each iteration at a fixed measurement setting $\tilde{u}$, which we take to be equal to the initial guess for the spin resonance. The twin uses Eq. \ref{eq:fisher-optimization} to select a set of optimal values for the drive frequency for which the measurement result is expected to depend most strongly on both the magnetic field $B_0$ and the Rabi frequency $\Omega$, given the particular $\phi(t)$ obtained for that iteration.

\begin{figure}
    \centering
    \includegraphics[width=0.4\linewidth]{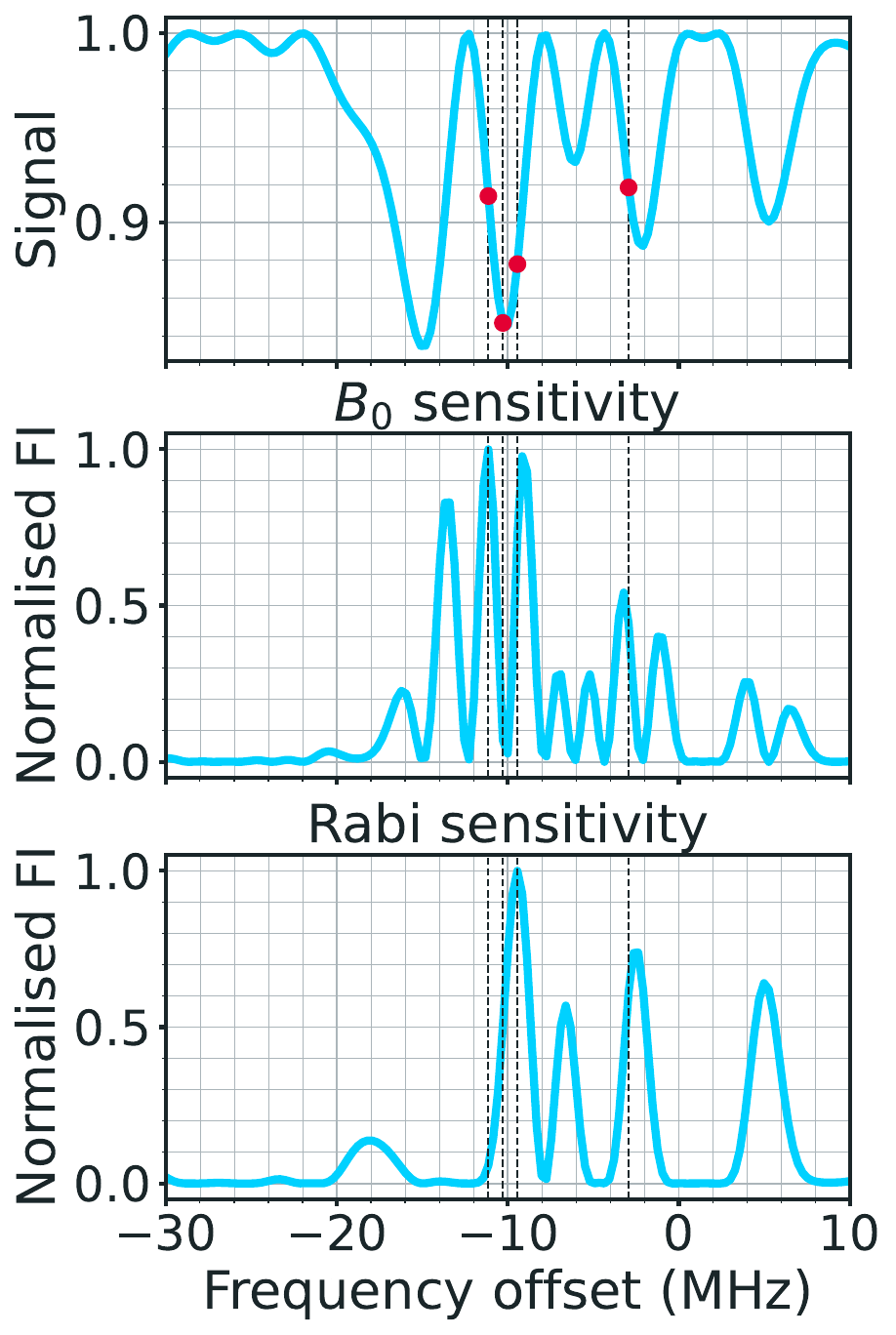}
    \caption{\textbf{Fisher-information validation for pulsed ODMR parameter estimation.}
Top: learned ODMR spectrum as a function of frequency offset, with Fisher-information-selected probe frequencies indicated by red markers and dashed vertical lines. Middle and bottom: normalized single-parameter Fisher information for the static field $B_0$ and the Rabi frequency $\Omega$, respectively. The selected probe frequencies concentrate near spectral regions that are informative for estimating these parameters, while the distinct sensitivity profiles illustrate their complementary frequency dependence.}
    \label{fig:podmr_fi_validation}
\end{figure}

\begin{figure}[htbp]
    \centering
    \includegraphics[width=0.4\linewidth]{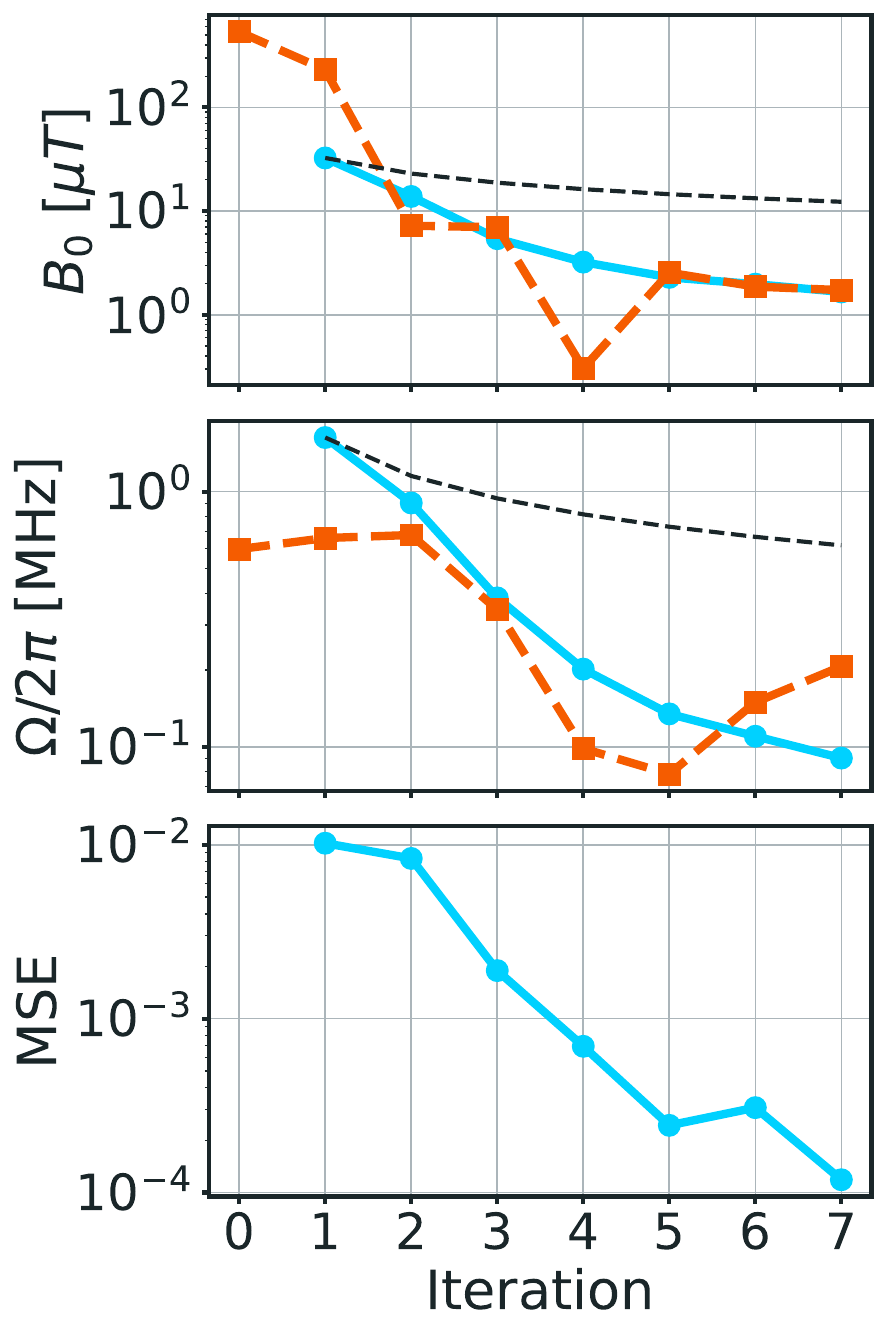}
    \caption{\textbf{Convergence and error quantification during iterative PODMR calibration.}
The first two panels show the estimated Gauss--Newton (GN) $1\sigma$ uncertainties (cyan circles) and absolute parameter-estimation errors (orange dashed squares) for the static magnetic field $B_0$ and the Rabi
frequency $\Omega/2\pi$, respectively. The orange point at iteration $0$ denotes the error of the initial model relative to the KB reference; GN uncertainties are available only after the first FI-selected measurement batch. The black dashed curves show the ideal $1/\sqrt{n}$ scaling expected as the accumulated number of measurements $n$ increases. The third panel reports the mean-squared error (MSE) between the measured FI-selected probe signals and the pre-fit model predictions for each iteration. Together, the panels quantify the reduction in parameter uncertainty, agreement with the KB reference, and improvement in model-to-measurement agreement throughout the calibration sequence.}
    \label{fig:podmr_error_evolution}
\end{figure}

Data from an example calibration run using ML1P can be seen in Figs. \ref{fig:podmr_fi_validation} and \ref{fig:podmr_error_evolution}, demonstrating fast convergence and excellent agreement between the final simulation and corresponding measurements. Fig. \ref{fig:podmr_fi_validation} is a visualization of the measurement setting selection process. Panel (a) shows the simulated response across the range of measurement settings for a given iteration - this is what the FI calculation is based on. Panels (b) and (c) show Eq. \ref{eq:scalar-fi} evaluated with respect to the magnetic field $B_0$ and the Rabi frequency $\Omega$ only. Together the three panels demonstrate how the measurement selection process selects for settings that are jointly informative with respect to the parameters to be learned, with points on the slope of the line-shape being informative of the magnetic field $B_0$ and points towards some of the maxima being informative of the Rabi frequency $\Omega$.

Figure \ref{fig:podmr_error_evolution} shows the convergence of the iterative learning loop. Panel (a) is the mean squared error (MSE) between the signal predicted by the digital twin at the FI selected measurement settings, and the corresponding real data returned from the QPU. It is important to point out that in this plot the MSE is calculated independently for each iteration before the learning step and is used only to measure the convergence of the model. This is distinct from the loss function defined in Eq. \ref{eq:learning-loss} which is evaluated on all batches of controls and measurements and used to update the model parameters during the learning step. After the feedback loop has converged the Gauss--Newton (GN) covariance method (Section \ref{sec:parameter-error-estimation}) is applied to the accumulated experimental data to attach $1\sigma$ standard error bars to the model parameters at every iteration step. The standard error together with the absolute difference to the parameter value as obtained from the normal calibration is shown in panels (b) and (c) of Fig. \ref{fig:podmr_error_evolution}. It is particularly interesting to note that the error reduction over time is significantly faster than what would be expected from a simple $\sqrt{n}$ scaling with measurement time. This demonstrates the power of iteratively improving the measurement settings in response to the model updates.

\begin{figure}
    \centering
    \includegraphics[width=0.7\linewidth]{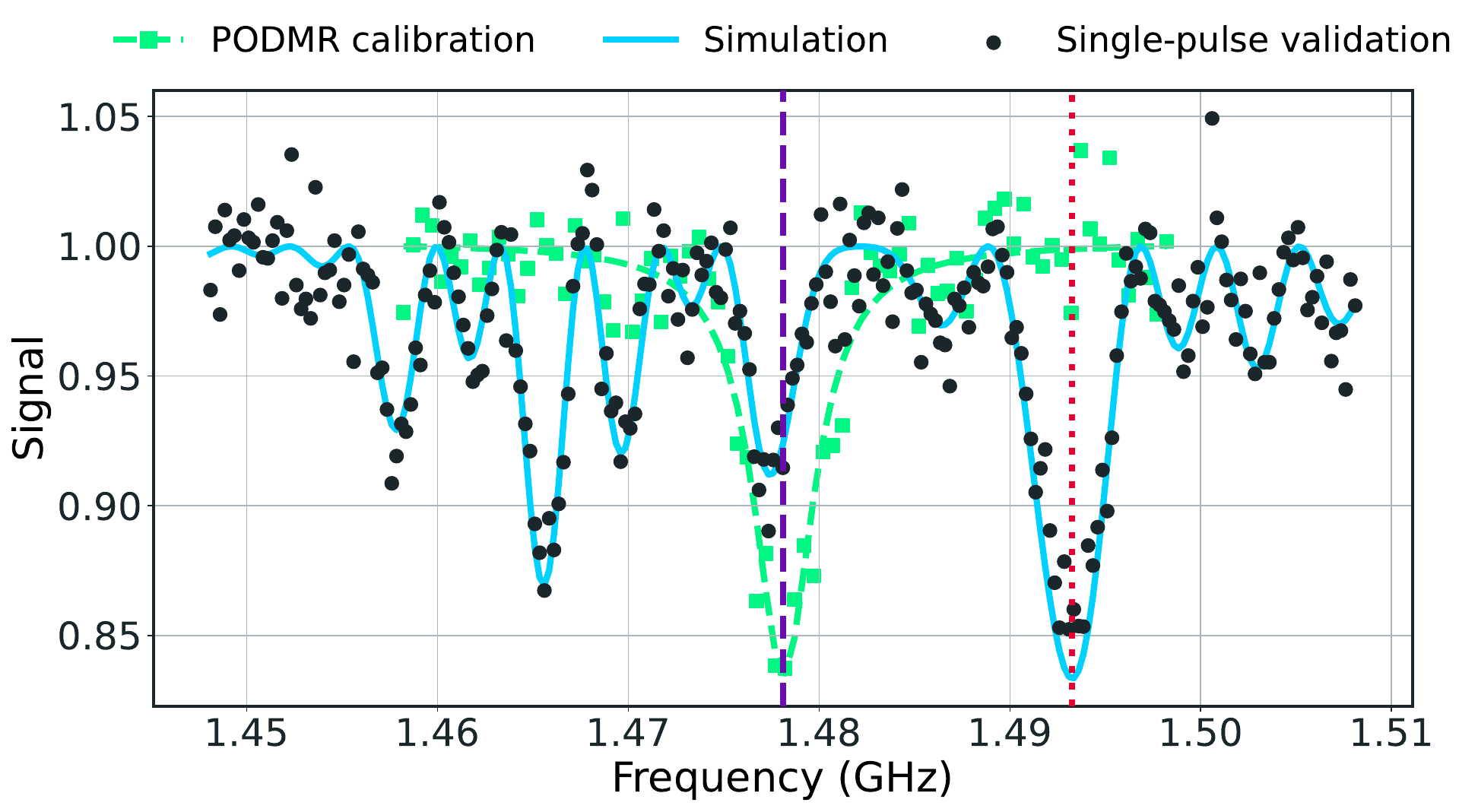}
    \caption{\textbf{Verification of the final model across the measurement setting space.} We applied the control pulse obtained during iteration three over a large regularly spaced grid of drive frequencies and also simulated the outcome using the digital twin with the final parameters obtained from the calibration.
    We also added the PODMR standard calibration points to the plot, along with the corresponding fit as a dashed line.
    The learned qubit frequency has been marked with a purple dashed vertical line in the plot at 1.47810 GHz, matching the overlaid conventional PODMR spectrum. The initial guess for the qubit frequency is marked with vertical dotted red line at 1.49326 GHz.}
    \label{fig:podmr-results}
\end{figure}

To validate the accuracy of the final learned model we took the control pulse obtained during the last iteration and applied it across a large range of measurement settings (rather than specific informative points only). The results, together with the corresponding model prediction, are shown in Fig. \ref{fig:podmr-results}. The excellent agreement between the data and the simulation across the full range of possible measurement settings
suggests that the model, which is used both to calculate the pulse shape and to predict the effect of the pulse, is accurately capturing the relevant physics of the experiment. The learned qubit frequency has been marked with a green vertical line in the plot, matching the overlaid conventional PODMR spectrum. The initial guess for the qubit frequency is marked with vertical dotted purple line. The measured contrast at this point is maximal, consistent with the fact that the pulse optimization objective is a complete population inversion at the initial guess for the frequency.
We highlight that this is a zero free parameter simulation, meaning the model parameters have not been updated to match the data.
Finally, it is interesting to note that the response of the phase modulated control pulse has a large bandwidth of more than 50 MHz despite a Rabi rotation rate for the NV electron of only 3 MHz. Note that the pulse was not optimized for having a uniform or broadband response, rather the fact that it was calculated to affect a $\pi$-inversion at a single off-resonant frequency results in a highly non-uniform response over a very large bandwidth as compared to the Rabi rotation rate. The broadband response helps the digital twin quickly pin down the qubit frequency, even when the initial guess is far off resonance. In the example shown, the initial guess for the qubit frequency was off by 15 MHz, and already after two iterations, with four measurement points each, the digital twin recovers the correct qubit frequency with an error of $300$ kHz, as seen in Fig. \ref{fig:podmr_error_evolution}.
This can be compared with the standard PODMR calibration shown in Figure \ref{fig:nv1-latest-calibrations}a, which had a standard error of $108\rm{kHz}$ after a uniform sweep of 81 measurement settings.
\subsection{Calibration method comparison}
We repeated the calibration workflow for the different experimental designs shown in the schematic in Fig. \ref{fig:sequences}. Table~\ref{tab:gn_parameter_errors} summarizes the learned parameter values, the Gauss--Newton $1\sigma$ error, and the measurement duration for the example sequences with detailed results available in the \teal{appendix [ref]}. For comparison, the table also includes the equivalent data for the conventional calibration sequences presented in \ref{subsec:experimental}.
We note that the GN error for both $\omega_q$ and $\Omega$, is lower for ML2P than ML1P and lower still for the ML4P. As expected, the addition of a free evolution period increases the available FI per measurement point, which together with the more varied learning landscape of a longer sequence lets the model pin the parameters with greater accuracy. Beyond accuracy, the addition of a free evolution period allows the model to learn new parameters that do not vary with pulse amplitude or duration alone, like the dephasing rate $1/T_2^\star = \gamma$ and the post-pulse delay time $\tau_p$. We added such a delay time manually as a proof of principle regarding the ability of the framework to learn unknown hardware calibration constants.

All three ML sequences finish in significantly shorter time than the full KB calibration, with comparable fit errors. The qubit frequency predicted by ML2P and ML4P is slightly lower (by 20-30 kHz) than that obtained by the standard calibration.
\\

\begin{table*}[t]
\centering
\small
\begin{tabular}{ccrrrrr}
\toprule
Experiment & Parameter & Measured value & Error & QPU Time\\
\midrule
PODMR
& $\omega_q$ [$\mathrm{kHz}$]
& 1477990.314 & 108 & 55 s \\
\midrule
Rabi
& $\Omega/2\pi$ [$\mathrm{MHz}$]
& 2.976 & 0.0132 & 240 s\\
\midrule
Ramsey chevrons
& $\omega_q$ [$\mathrm{kHz}$]
& 1478151.055 & 1.748 & 811 s\\
& $\gamma$ [$\mathrm{kHz}$]
& 319 & 5.397\\
\midrule
Combined KB Calibration
& $\omega_q$ [$\mathrm{kHz}$]
& 1478151.055 & 1.748 & $\boldsymbol{1106 \mathrm{s}}$\\
& $\Omega/2\pi$ [$\mathrm{MHz}$]
& 2.976 & 0.0132\\
& $\gamma$ [$\mathrm{kHz}$]
& 319 & 5.397\\
\midrule
ML1P
& $\omega_q$ [$\mathrm{kHz}$]
& 1478109.316 & 46.5 & 38 s\\
& $\Omega/2\pi$ [$\mathrm{MHz}$]
& 3.1828 & 0.0904 \\
\midrule
ML2P
& $\omega_q$ [$\mathrm{kHz}$]
& 1478127.941 & 11.9 & 220 s\\
& $\Omega/2\pi$ [$\mathrm{MHz}$]
& 2.9912 & 0.0273 \\
& $\tau_p$ [$\mathrm{ns}$]
& 99.4205 & 1.6945 \\
& $\gamma$ [$\mathrm{kHz}$]
& 350.7249 & 45.6093 \\
\midrule
ML4P
& $\omega_q$ [$\mathrm{kHz}$]
& 1478118.094 & 3.4 & 275 s\\
& $\Omega/2\pi$ [$\mathrm{MHz}$]
& 3.0038 & 0.0181 \\
\bottomrule
\end{tabular}
\caption{Calibration results by method. The error is the standard error of the analytical fit for the the three conventional calibration methods. For the ML methods the error is the GN error as described in \ref{sec:parameter-error-estimation}}
\label{tab:gn_parameter_errors}
\end{table*}

\subsection{Validation/Benchmarking}
\label{sec:benchmarking}
To assess the quality of the model-based calibrated parameters relative to those obtained from standard fitting procedures, we consider quantum process tomography (QPT) and randomized benchmarking (RB) as independent validation steps. These benchmarks test whether the pulses optimized from the model-based calibration are better suited to the current experiment than pulses derived from conventional calibration methods, by comparing the resulting gate performance under both parameter sets.

\paragraph{QPT}
Quantum process tomography (QPT) provides an experimental, model-independent characterization of the gates implemented with the calibrated parameters, by reconstructing the process matrix from a tomographically complete set of input states and measurement bases~\cite{nielsen2000QuantumComputationInformation}.
We perform single-qubit QPT on the NV electron spin using 6 input states ($\pm Z$, $\pm X$, $\pm Y$) and 3 readout axes ($Z$, $X$, $Y$), applied for 3 pulse angles (identity, $\pi/2$ gate, and $\pi$ gate), giving 54 measurement settings in total.
Each setting is accumulated with $N=200{,}000$ shots on the NV1 qubit of the XeedQ QPU (XQ1), with $\pi$-pulse duration $t_\pi = 168\,\mathrm{ns}$ and nominal Rabi frequency $\Omega_0 \approx 2\pi \times 2.976\,\mathrm{MHz}$. The process matrix is estimated by maximum-likelihood estimation (MLE) constrained to completely-positive trace-preserving (CPTP) maps; the theoretical background, including the Choi--$\chi$ relation and the fidelity metrics, is detailed in Appendix~\ref{sec:qpt-theory}.\\
Figure~\ref{fig:qpt-identity-pi2-pi} shows the experimental infidelities for  all three pulse angles side by side, comparing standard calibration against ML-based calibration from the four and one pulse experiments, together with the corresponding process fidelity and gate fidelity for each method.

\begin{figure}[htbp]
    \centering
    \includegraphics[width=1.0\linewidth]{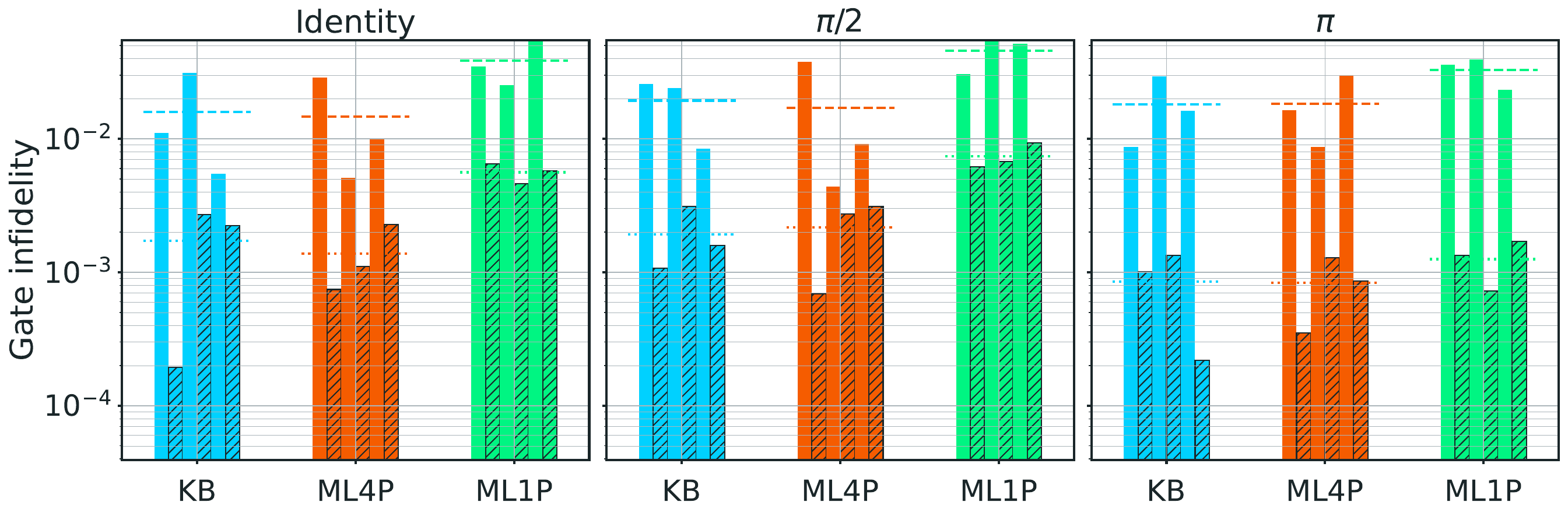}
    \caption{\textbf{Total and coherent gate infidelity from QPT measurements.}
Gate infidelities are shown on a logarithmic scale for identity, $\pi/2$, and
$\pi$ gates (left to right) across the KB, four-pulse, and one-pulse
calibration protocols. For each calibration and each of three repetitions, averaged over $200000$ measurements,
solid bars denote the total Choi infidelity,
$1-F_{\mathrm{g}}(\mathcal{E}_{\mathrm{Choi}})$, whereas hatched bars denote
the coherent infidelity of the closest unitary channel,
$1-F_{\mathrm{g}}(U_{\mathrm{closest}})$. Dashed and dotted horizontal
segments indicate the corresponding repetition-averaged Choi and
closest-unitary infidelities, respectively. The difference between the solid
and hatched values quantifies the incoherent contribution, including
decoherence and SPAM effects, which is not removed by coherent recalibration.
}
    \label{fig:qpt-identity-pi2-pi}
\end{figure}

\begin{figure}[htbp]
  \centering
  \includegraphics[width=0.7\linewidth]{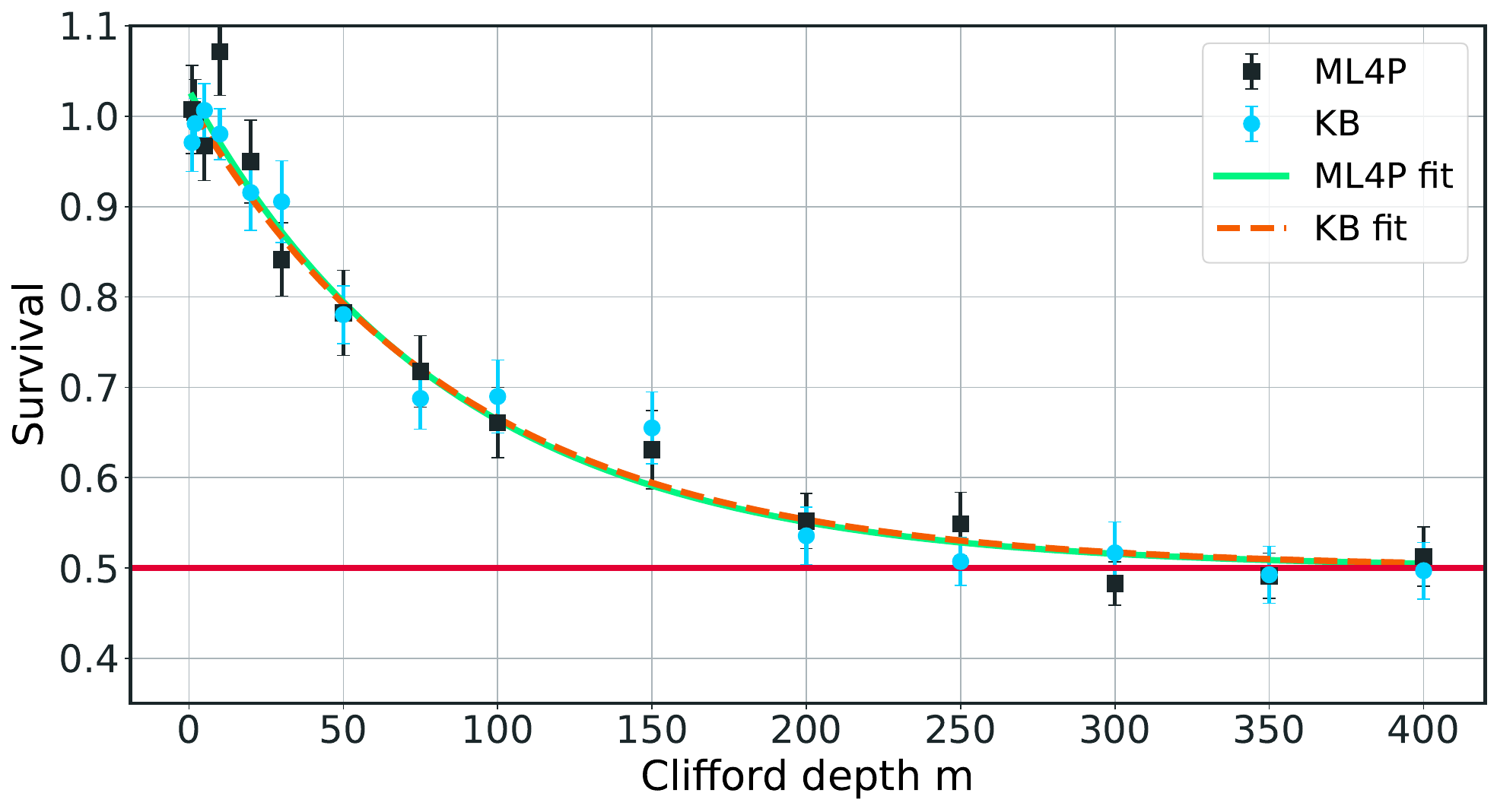}
  \caption{Single-qubit RB survival-probability curves for KB and ML4P learned parameteres. The post-correction fit fixes the SPAM parameters to values obtained by offline model learning of selected RB curve points using the digital twin, isolating the intrinsic gate depolarisation rate $\alpha$ and yielding to the error per Clifford $\mathrm{EPC}=\tfrac{1}{2}(1-\alpha)$. From the fit, gate fidelities are $F_g = 99.4178\%$ for ML4P and $F_g = 99.4377\%$ in the case of KB.}
  \label{fig:rb-corrected}
\end{figure}

\paragraph{RB}
As a second independent characterization of the implemented gates fidelity, a single-qubit randomized benchmarking (RB)~\cite{magesan2011ScalableRobustRandomized} was performed providing an independent estimate of the average error per Clifford gate. The RB measurement was performed up to a Clifford depth of 400 and with 30 sequences per Clifford depth. 
Figure~\ref{fig:rb-corrected} shows the RB survival-probability curve after SPAM correction (refer to Appendix \ref{sec:rb_analysis} for more details about the correction analysis). The full depolarising model $f(m) = A\,\alpha^m + b$ fit is also shown in the figure, leading to the depolarisation rate $\alpha$ and yielding the error per Clifford $\mathrm{EPC}=\tfrac{1}{2}(1-\alpha)$.

\section{Conclusion}
\label{sec:conclusions}
We have presented a unified closed-loop model-based calibration workflow applied to NV-centre quantum processing units that addresses the measurement-efficiency and accuracy limitations of conventional sequential sweep-and-fit routines.
The framework combines three components: a physics-based digital twin of the system, a Fisher-information-guided experiment design that selects the most informative probe conditions at each iteration, and a gradient-descent model-learning step that refits the Hamiltonian parameters to the accumulated experimental record.
Together, these components form a self-improving loop in which both the control pulse and the measurement conditions grow more informative as the parameter estimates converge, enabling faster-than-$\sqrt{N}$ scaling of estimation accuracy with QPU time.

Across the four pulse sequences demonstrated here, FI-guided probe selection consistently outperformed heuristic or hardcoded grids, requiring fewer QPU calls to reach the same parameter accuracy.
The two-pulse sequence experiment in particular illustrates the efficiency gain: the FI-guided loop converges in a handful of iterations with a small number of carefully chosen points, whereas a conventional dense measurement approach requires more than five times more QPU calls to cover the same control space.
The four-pulse sequence, which has no direct conventional counterpart due to its non-analytical signal shape, was made tractable precisely because the FI-guided design avoids uninformative regions of the two-dimensional control space.

The framework additionally generalises naturally from calibration to characterization: the XY8 dynamical-decoupling experiment recovers both the parallel and transverse hyperfine couplings of a weakly coupled $^{13}$C spin to high accuracy, demonstrating that the same loop architecture transfers across qualitatively different experimental regimes without modification.

Independent gate-level validation using both QPT and single qubit RB on the XeedQ NV1 QPU confirms that the model-based calibration produces control pulses whose quality is statistically equivalent to those obtained from standard sweep-and-fit procedures, while requiring fewer QPU measurements.
Quantum process tomography across three pulse angles (identity, $\pi/2$, and $\pi$) yields comparable process and average gate fidelities between the ML-based and conventional calibration methods for both the multi-tau Ramsey and PODMR sequences. Likewise, single-qubit randomized benchmarking with offline SPAM correction shows that the per-Clifford error rates of the two approaches are indistinguishable within experimental uncertainty.

The current implementation demonstrates that model-based calibration is competitive with standard techniques even on hardware limited by run-to-run drift, where individual gate infidelity runs cannot statistically resolve the two methods.
Whenever the underlying simulation offers a good description of the quantum hardware, the loop architecture is platform agnostic and scales favourably as the number of parameters grows: whereas conventional sequential sweeps require a number of measurements that scales at least linearly with the number of parameters, the FI-guided loop amortises each measurement across all parameters simultaneously through the joint Fisher information matrix.
Natural extensions include multi-qubit calibration where cross-talk and two-qubit interaction parameters must be learned jointly, real-time adaptive probe reselection within each iteration, and co-optimisation of the control pulse and the probe design under a shared budget constraint.

\section*{Acknowledgments}
We would like to thank Benjamin Habie and Roman Razilov for help with the IT infrastructure and the Qruise team for helpful discussions.
\appendix

\section{Fisher information model}
\label{app:fisher-information-model}
This appendix derives the fisher-information matrix responsible for the choice of the most informative measurement settings.\\
The digital twins use a  the photon-counting readout model. Let $p \equiv p(\boldsymbol{u},\boldsymbol{\theta},\boldsymbol{c}^{\star})$ denotes the simulated population in the dark $m_s=-1$ state. The measured signal is
\begin{equation}
s = \mathcal{C}_{\mathrm{ODMR}}\,p,\qquad \mathcal{C}_{\mathrm{ODMR}} =\frac{\lambda_{m_s=0}-\lambda_{m_s=-1}}{\lambda_{m_s=0}},
\end{equation}
where $\lambda_{m_s=0}$ and $\lambda_{m_s=-1}$ are the mean photon counts per readout window for the bright and dark states, respectively. The population-equivalent variance used by the digital twin is
\begin{equation}
V_p(p) = \frac{(1-p)\lambda_{m_s=0}+p\lambda_{m_s=-1}+\left(\lambda_{m_s=0}-\lambda_{m_s=-1}\right)^2 p(1-p)}{\left(\lambda_{m_s=0}-\lambda_{m_s=-1}\right)^2},
\label{eq:podmr-readout-variance}
\end{equation}
with a numerical lower bound $V_p \geq 10^{-12}$. The first two terms describe Poisson fluctuations of the bright and dark fluorescence counts, whereas the last term is the binomial population-assignment contribution.\\
Accordingly, the Fisher-information-matrix coefficients evaluated by the digital twins are
\begin{equation}
\mathcal{I}_{ab}(\boldsymbol{u},\boldsymbol{\theta},\boldsymbol{c}^{\star})=\frac{1}{V_p(p)}\frac{\partial s}{\partial\theta_a}\frac{\partial s}{\partial\theta_b}=\frac{\mathcal{C}_{\mathrm{ODMR}}^2}{V_p(p)}\frac{\partial p}{\partial\theta_a}\frac{\partial p}{\partial\theta_b},
\label{eq:scalar-fi-detailed-model}
\end{equation}
where $(a,b)$ index the model parameters. 

\section{Greedy D-Optimal Measurement Selection}
\label{app:doptimal}

This appendix provides a concise mathematical account of the greedy D-optimal selection algorithm used in Section~\ref{sec:fi-selection} to choose the set of measurement settings $\{\boldsymbol{u}^{\star}\}$ at each calibration iteration.

\paragraph{D-optimality criterion.}
The Cram\'{e}r-Rao bound states that the covariance of any unbiased estimator $\hat{\boldsymbol{\theta}}$ satisfies
\begin{equation}
  \mathrm{Cov}(\hat{\boldsymbol{\theta}})
  \;\geq\;
  \left(\sum_{i=1}^{M} \mathcal{I}(\boldsymbol{u}_i,\boldsymbol{\theta},
  \boldsymbol{c}^{\star})\right)^{-1},
\end{equation}
where $\mathcal{I}_{ab}$ is the FIM defined in Eq.~\eqref{eq:scalar-fi}. The volume of the joint confidence ellipsoid for $\boldsymbol{\theta}$ is inversely proportional to $\sqrt{\det(\sum_i \mathcal{I}_i)}$, so the design score $S$ in Eq.~\eqref{eq:fisher-optimization}---maximising $\det(\sum_i \mathcal{I}_i)$---directly minimises this volume~\cite{kiefer1960equivalence,pukelsheim1993optimal}. This is the classical \emph{D-optimality} criterion.

\paragraph{Rank-1 structure and the Matrix Determinant Lemma.}
For binomial readout, every individual setting $k$ contributes a rank-1 matrix
\begin{equation}
  \mathcal{I}_k
  = \frac{\mathbf{g}_k \mathbf{g}_k^T}{\sigma_k^2},
  \qquad
  \mathbf{g}_k = \nabla_{\boldsymbol{\theta}}\, p(\boldsymbol{u}_k),
  \quad
  \sigma_k^2 = p_k(1-p_k),
\end{equation}
where the gradients are obtained via automatic differentiation through the digital twin. Let $\mathbf{F} = \sum_{j<k} \mathcal{I}_j$ be the accumulated FIM from settings already selected.
By the Matrix Determinant Lemma~\cite{harville1997matrix},
\begin{equation}
  \det\!\bigl(\mathbf{F} + \mathcal{I}_k\bigr)
  = \det(\mathbf{F})
    \Bigl(1 + \mathbf{g}_k^T \mathbf{F}^{-1} \mathbf{g}_k\,/\,\sigma_k^2\Bigr),
  \label{eq:app-mdl}
\end{equation}
so adding a new point multiplies the determinant by a factor equal to one plus its \emph{leverage} $d_k^2 = \mathbf{g}_k^T \mathbf{F}^{-1} \mathbf{g}_k / \sigma_k^2$.  The optimal next setting is therefore the candidate with the highest leverage under the current accumulated information.

\paragraph{Greedy sequential algorithm.}
Because searching all $\binom{M_{\mathrm{grid}}}{M}$ subsets is NP-hard, we use a greedy scheme. Since any single-point FIM is rank-1 and has zero determinant for$D>1$, the accumulated matrix is initialised with a small regulariser $\mathbf{F}^{(0)} = \varepsilon\mathbf{I}_D$ ($\varepsilon = 10^{-12}$) to ensure positive-definiteness throughout the loop.
At each step $k = 1, \ldots, M$:
\begin{enumerate}
  \item For every candidate $(\boldsymbol{u}_{ij})$ on the 2-D grid, form
        $\mathbf{F}_{\mathrm{cand}} = \mathbf{F}^{(k-1)} + \mathcal{I}_{ij}$.
  \item Evaluate the score via the \emph{log-determinant}
        $(\mathrm{sign},\,\ell) = \mathrm{slogdet}(\mathbf{F}_{\mathrm{cand}})$
        for numerical stability (avoids overflow / underflow in
        $\det$)~\cite{pukelsheim1993optimal}.
        Set $\mathrm{score} = \ell$ if $\mathrm{sign}>0$, else $-\infty$.
  \item Select $\boldsymbol{u}_k^{\star} = \arg\max_{ij}\,\mathrm{score}$ and
        update $\mathbf{F}^{(k)} = \mathbf{F}^{(k-1)} + \mathcal{I}_{k}$.
  \item Mask out a neighbourhood of $\boldsymbol{u}_k^{\star}$ to enforce a
        minimum physical separation between selected settings.
\end{enumerate}
\begin{figure}
    \centering
    \includegraphics[width=0.7\linewidth]{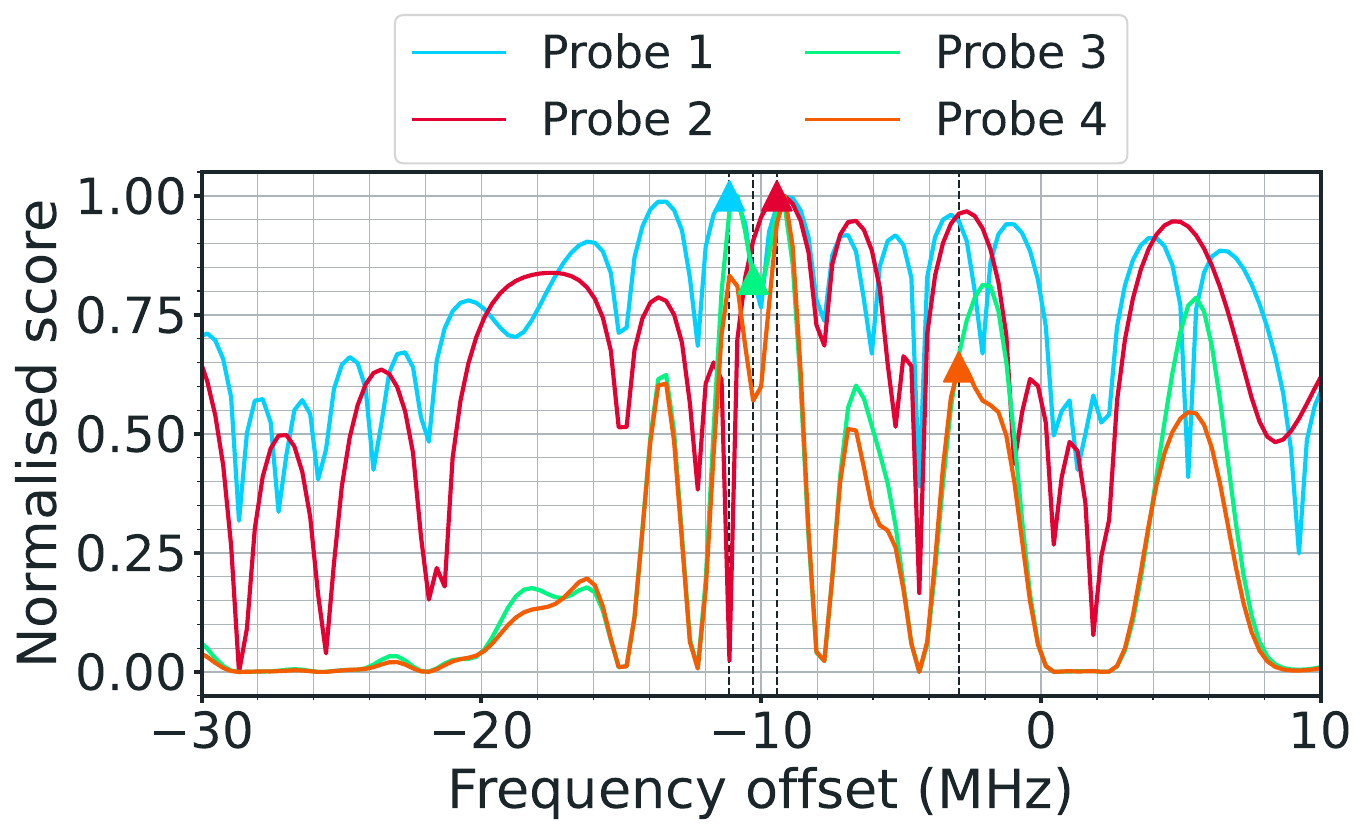}
    \caption{\textbf{Greedy D-optimal probe-selection scores for pulsed ODMR.}
Normalized candidate-frequency score curves are shown after each of four sequential probe selections. The curves quantify the incremental information gain associated with adding a probe at each frequency offset, conditioned on probes selected previously. Triangular markers and dashed vertical lines identify the selected frequencies for probes 1--4. The selected probes target complementary high-information regions of the ODMR spectrum to improve joint parameter estimation.}
    \label{fig:podmr-greedy-score}
\end{figure}

\paragraph{Approximation guarantee.}
The objective $f(\mathcal{X}) = \log\det\bigl(\sum_{k\in\mathcal{X}} \mathcal{I}_k\bigr)$ is \emph{submodular} and monotonically non-decreasing~\cite{nemhauser1978analysis,krause2005near}.
By Nemhauser, Wolsey and Fisher's theorem~\cite{nemhauser1978analysis}, greedy maximisation of a monotone submodular function yields a set $\mathcal{X}_{\mathrm{greedy}}$ satisfying
\begin{equation}
  f(\mathcal{X}_{\mathrm{greedy}})
  \;\geq\;
  \Bigl(1 - e^{-1}\Bigr)\, f(\mathcal{X}_{\mathrm{opt}})
  \;\approx\;
  0.63\, f(\mathcal{X}_{\mathrm{opt}}),
\end{equation}
and the total computational cost is $\mathcal{O}(M \cdot M_{\mathrm{grid}})$ instead of exponential.

\section{Qubit Transition Frequency from the Longitudinal Field}

The digital twin uses the longitudinal static magnetic field $B_0$ to determine the resonance frequency of the selected NV-centre transition.
$B_0$ is specified in Tesla, whereas all quoted transition frequencies and Hamiltonian parameters are expressed in Hz.\\
For the secular two-level interpretation of the chosen electronic-spin and $^{14}\mathrm{N}$ nuclear-spin transition, the resonance frequency is approximated by
\begin{equation}
    f_q(B_0)=D- \gamma_e B_0- A_{\parallel}^{(^{14}\mathrm{N})},
    \label{eq:qubit-frequency-secular}
\end{equation}
where $D$ is the NV zero-field splitting, $\gamma_e$ is the electron gyromagnetic ratio in Hz/T, and $A_{\parallel}^{(^{14}\mathrm{N})}$ is the parallel NV--$^{14}\mathrm{N}$ hyperfine coupling for the selected nuclear spin projection.

The default parameters used by the digital-twin problem factory are
\begin{align}
    D &= 2.87~\mathrm{GHz}, \\
    \gamma_e &= -2.802 \times 10^{10}~\mathrm{Hz/T}, \\
    A_{\parallel}^{(^{14}\mathrm{N})} &= -2.14~\mathrm{MHz}.
\end{align}
Consequently, for positive $B_0$, the term $-\gamma_e B_0$ increases the transition frequency under the convention used by the twin.\\
For a small field update $\delta B_0$, the secular approximation gives the linear sensitivity
\begin{equation}
    \delta f_q
    =
    -\gamma_e \, \delta B_0,
    \label{eq:qubit-frequency-sensitivity}
\end{equation}
so that
\begin{equation}
    \left| \frac{\partial f_q}{\partial B_0} \right|
    =
    |\gamma_e|
    \approx 28.02~\mathrm{GHz/T}
    =
    28.02~\mathrm{kHz/\mu T}.
\end{equation}
This relation is used to interpret a learned $B_0$ correction as a corresponding shift of the microwave resonance.\\
The reported qubit frequency is calculated more generally than Eq.~\eqref{eq:qubit-frequency-secular}.  After defining the full static Hamiltonian in terms of the static magnetic field $B_{0}$,
\begin{equation}
    H_\mathrm{static}(\hat{B}_0),
\end{equation}
including the zero-field splitting, electronic and nuclear Zeeman terms, hyperfine couplings, quadrupole terms, and any enabled non-secular terms, we diagonalizes it as :
\begin{equation}
    H_\mathrm{static}(\hat{B}_0)\ket{E_j}
    =
    E_j\ket{E_j}.
\end{equation}
The requested computational-basis states, denoted $\ket{\psi_\mathrm{initial}}$ and $\ket{\psi_\mathrm{target}}$, are matched to the eigenstates with maximum overlap:
\begin{align}
    j_\mathrm{i}
    &=
    \arg\max_j
    \left|
        \braket{E_j | \psi_\mathrm{initial}}
    \right|^2, \\
    j_\mathrm{t}
    &=
    \arg\max_j
    \left|
        \braket{E_j | \psi_\mathrm{target}}
    \right|^2.
\end{align}
Finally, the qubit transition frequency is
\begin{equation}
    \omega_q(\hat{B}_0)
    =
    \frac{
        \left|
            E_{j_\mathrm{t}} - E_{j_\mathrm{i}}
        \right|
    }{2\pi}.
    \label{eq:qubit-frequency-exact}
\end{equation}

\section{Other sequences application and results}
\label{Appendix:other_sequences}
\begin{figure}[h]
    \centering
    \begin{subfigure}[b]{1.02\linewidth}
        \centering
        \includegraphics[width=\linewidth]{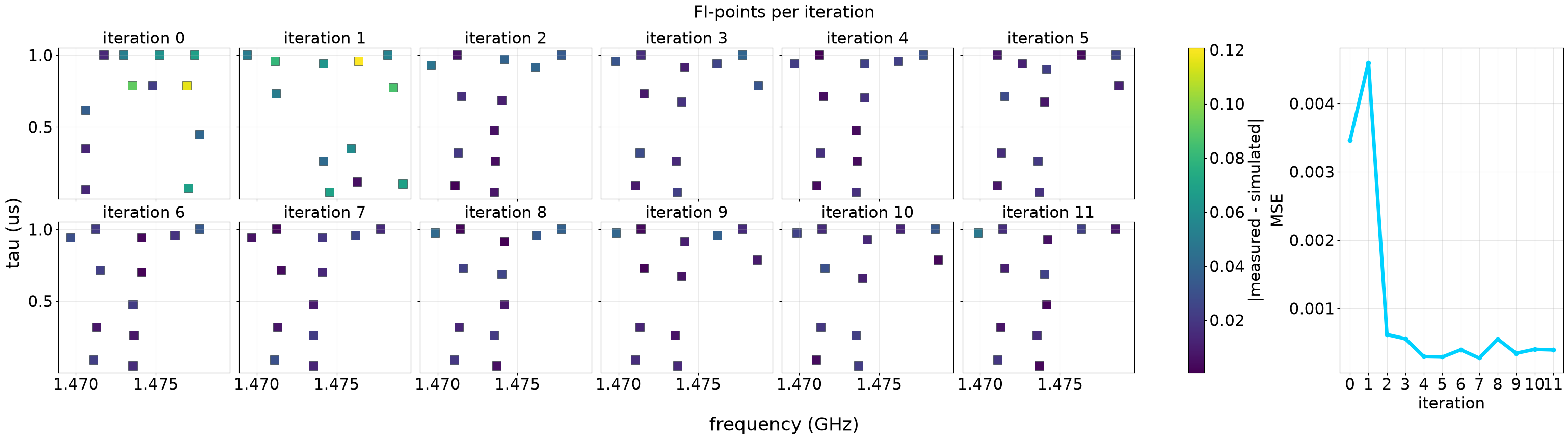}
        \caption{}
        \label{fig:twopulse_fi}
    \end{subfigure}

    \vspace{0.5em}

    \begin{subfigure}[b]{0.8\linewidth}
        \centering
        \includegraphics[width=\linewidth]{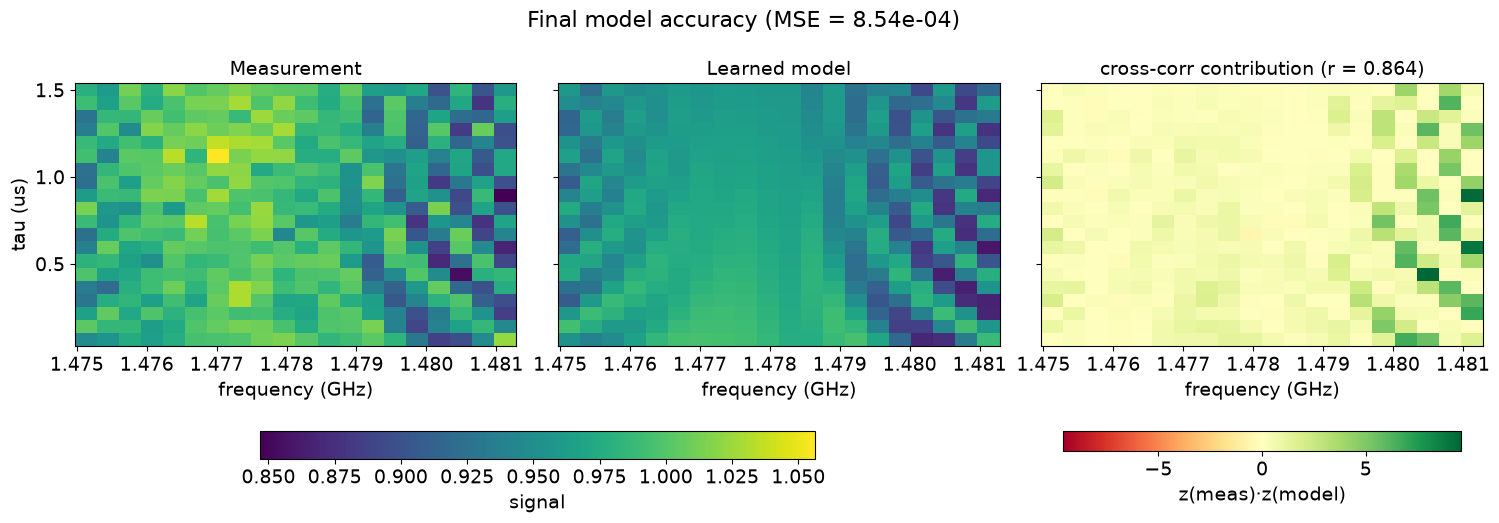}
        \caption{}
        \label{fig:twopulse_modelvsmeas}
    \end{subfigure}
    \caption{ML2P results (a) FI points taken over 11 iterations. Each point in a particular combination of $\tau$ and frequency. The color scale shows the residuals of the measured value and the simulated value for each iteration after the QOC step before the ML step. Right plot shows the MSE of the simulation pre-learning and the measured FI data points. (b) Left panel shows a dense grid measurement compared with the central panel that shows the equally dense grid digital twin results for the final values of the calibration workflow. Right panel shows the 2D cross-correlation per pixel comparing measurement and simulation \cite{Gonzlez2009DigitalIP}. Pearson correlation coefficient value $r$ shows good agreement.}
    \label{fig:smartramsey_combined}
\end{figure}

\paragraph{ML2P}
This experiment extends ML1P from one pulse to two, with a variable spacing between them, further increases the sensitivity of the sequence to the qubit resonance frequency, while also adding sensitivity to the dephasing time $T_2^*$. In this case there is richer 2D space of measurement settings, made up by the drive frequency and the spacing $\tau$ between the pulses. For this sequence we also inserted an artificial delay $\tau_p$ of 100ns after each microwave pulse and tasked the model with recovering it from an initial guess of 10ns. The total number of parameters to learn was thus 4.
Figure \ref{fig:smartramsey_combined} shows data from a calibration run using this sequence. Panel (a) shows the selected measurement points in the 2D space of available measurement settings for each iteration of the loop. The color scale is the difference between the digital twin predicted measurement result at that value, and the measured result from the QPU. It is clear that the learned model stabilizes after the third iteration, the FI selection algorithm returns a very similar set of points for all iterations following the third one. This convergence can also be seen in the MSE plot to the right, where the MSE between the digital twin predicted values before the measurement, and the measurement result is plotted versus iteration. The plot shows clear convergence at the third iteration. Panel (b) shows a model validation plot similar to \ref{fig:podmr-results}. We have taken the shaped pulse calculated during the last iteration of the loop and simulated the effect over a uniform grid of available measurement settings with the final learned model parameters. We also performed the corresponding measurement on the QPU and the two plots are presented side by side. Note that just as in \ref{fig:podmr-results}, this is a zero-free parameter simulation - the 4 learned model parameters are not updated after the final iteration of the calibration loop.

\paragraph{ML4P}
The purpose of this experiment design was to reduce the periodicity of the measured signal as seen in Fig. \ref{fig:smartramsey_combined} and simplify finding a global parameter minimum. The pulse sequence consists of four evenly spaced microwave pulses with the space of measurement settings being the frequency of the pulses, $\omega$ and the spacing between them, $\tau$. In this case we tried to learn 2 parameters, $\omega_q$ and $\Omega$. $T_2^*$ was not recoverable from this experiment due to an apparent refocusing effect of the regular pulses (regardless of the pulse angle and phase). We saw no measurable decoherence even over 10+$\mu s$

For this sequence we used two different objectives for the pulse shaping, resulting in two very different pulses. Objective-1 optimized the return of the system to the initial state after 4 consecutive pulses - the resulting pulse favored by the optimizer had an effective rotation angle of 75 degrees about the axis e, tilted from the z axis by 5 degrees. Objective-2 asked for the phase envelope that would turn the pulse into a $\pi/2$ X-gate. As for the one and two pulse cases, for both objectives the microwave frequency was kept fixed at the initial guess for the spin resonance frequency during the optimization. 

\begin{figure}[h]
    \centering
    \begin{subfigure}[b]{1.02\linewidth}
        \centering
        \includegraphics[width=\linewidth]{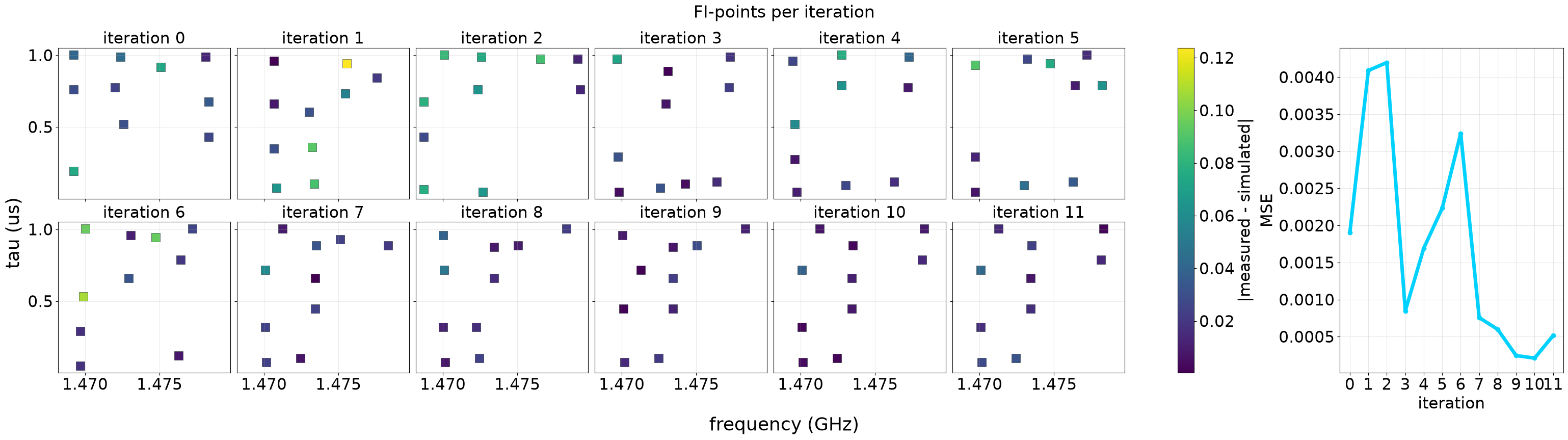}
        \caption{}
        \label{fig:fourpulse_fi}
    \end{subfigure}

    \vspace{0.5em}

    \begin{subfigure}[b]{0.8\linewidth}
        \centering
        \includegraphics[width=\linewidth]{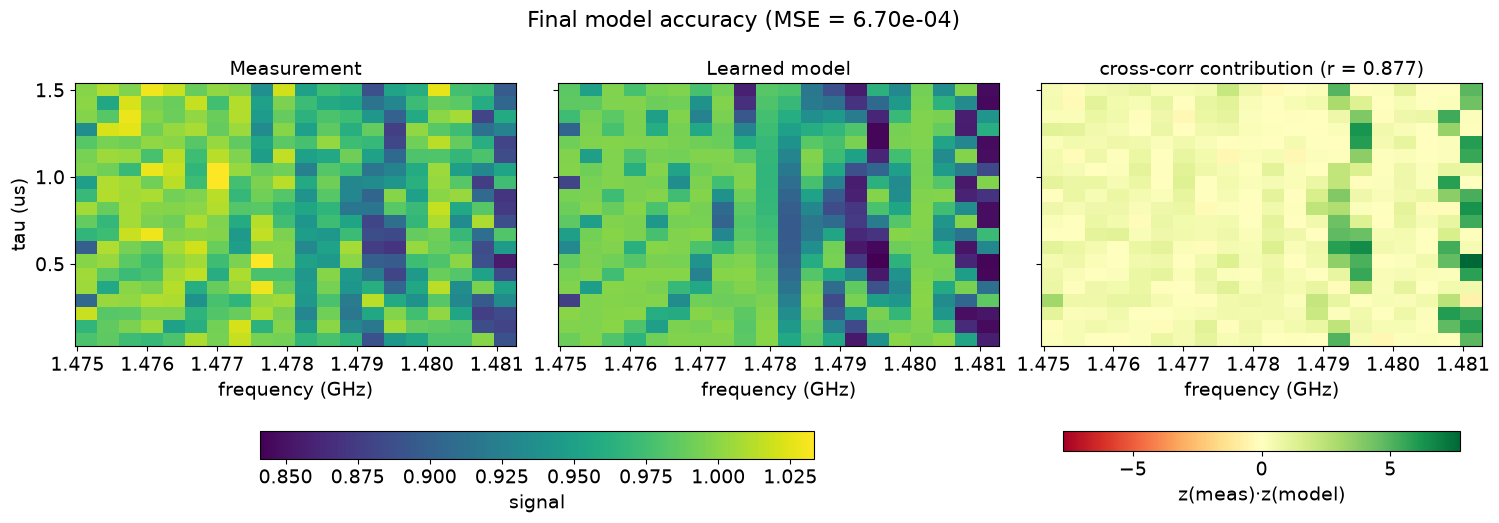}
        \caption{}
        \label{fig:fourpulse_modelvsmeas}
    \end{subfigure}
    \caption{ML4P sequence results (a) FI points taken over 11 iterations. Each point in a particular combination of $\tau$ and frequency. The color scale shows the residuals of the measured value and the simulated value for each iteration after the QOC step before the ML step. Right plot shows the MSE of the simulation pre-learning and the measured FI data points. (b) Left panel shows a dense grid measurement compared with the central panel that shows the equally dense grid digital twin results for the final values of the calibration workflow. Right panel shows the 2D cross-correlation per pixel comparing measurement and simulation \cite{Gonzlez2009DigitalIP}. Pearson correlation coefficient value $r$ shows good agreement.}
    \label{fig:multitau_combined}
\end{figure}

Results for a particular calibration run using ML4P are shown in Fig. \ref{fig:multitau_combined}. In these data we used pulses derived with the return to initial state objective. We see that it takes longer than ML2P to converge, but the final results are more accuracte as measured by the GN error as shown in \ref{tab:gn_parameter_errors}.

\paragraph{Dynamical Decoupling}
Dynamical decoupling spectroscopy, using the XY8 pulse sequence~\cite{delange2010UniversalDynamicalDecoupling,zhao2012AtomicScaleMultipartite}, is a well established method for identifying unresolved carbon-13 spins in the lattice surrounding a given NV center. In these experiments a periodic series of phase shifted pi pulses is applied to the NV spin at several different values for the inter pulse spacing, $\tau$. When $\tau$ is resonant with  

In XY8 dynamical-decoupling experiments, the inter-pulse spacing $\tau$ determines where resonance dips appear in the measured signal. 
The dip position is sensitive to the parallel hyperfine coupling $A_{\mathrm{C13},\parallel}$, while the dip shape and depth encode the perpendicular coupling $A_{\mathrm{C13},\perp}$.

\begin{figure}[h!]
    \centering
    \includegraphics[width=0.5\linewidth]{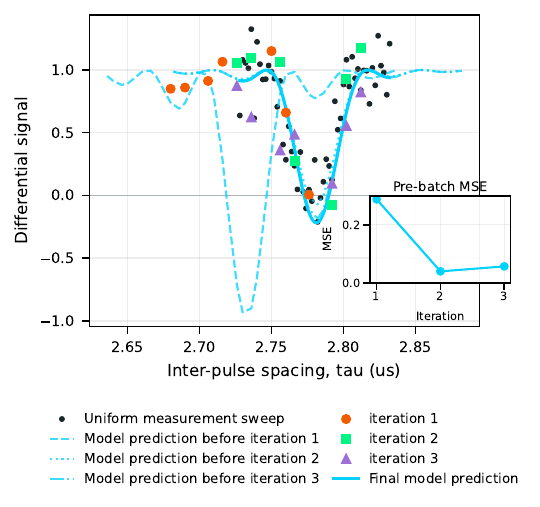}
    \caption{Dynamical Decoupling experiment, meant to obtain the hyperfine coupling parameters for a particular $^{13}$C in the diamond lattice. Overlay of the spectrum predicted by the fitted model with data obtained from a regular tau sweep.}
    \label{fig:hf_sensing}
\end{figure}

In Figure~\ref{fig:hf_sensing}, we can see a comparison between the predicted spectrum and the data obtained from a regular measurement for the dynamical decoupling experiment.

\section{Quantum Process Tomography machinery}
\label{sec:qpt-theory}

The underlying physical model is a 9-dimensional NV$+^{14}$N spin system
(NV electron $S=1$ coupled to N14 nucleus $I=1$).
The full unitary propagator $U(T) \in \mathbb{C}^{9\times 9}$ is computed from
the digital-twin Hamiltonian in the rotating frame; the N14 nuclear spin is then
traced out and the result projected onto the computational qubit subspace
$\{|m_s=0\rangle,\,|m_s=-1\rangle\}$ to obtain the effective $4\times 4$ Choi
matrix $\mathcal{C}$.

The Choi matrix is related to the $\chi$ process matrix in the normalised Pauli
basis $\{\hat{B}_m\} = \{I/\sqrt{2},\,\sigma_x/\sqrt{2},\,\sigma_y/\sqrt{2},\,
\sigma_z/\sqrt{2}\}$ by a unitary change of basis,
\begin{equation}
  \chi = \frac{1}{d}\,\Omega^\dagger\,\mathcal{C}\,\Omega,
  \qquad
  [\Omega]_{ki+m} = [\hat{B}_m]_{k,i},
  \label{eq:chi-from-choi}
\end{equation}
with $\mathrm{Tr}[\chi]=1$ for a trace-preserving channel.
For an ideal $X$ gate ($\pi$ pulse) the single dominant element is $\chi_{11}\approx 1$ (pure $\sigma_x$ rotation), with all other entries vanishing.
The process fidelity and average gate fidelity are
\begin{equation}
  F_{\mathrm{proc}}
    = \frac{\mathrm{Tr}[\mathcal{C}_{U^\dagger}\mathcal{C}]}{d^2},
  \qquad
  F_{\mathrm{avg}}
    = \frac{d\,F_{\mathrm{proc}}+1}{d+1},
  \label{eq:fidelity-metrics}
\end{equation}
so that $1-F_{\mathrm{avg}} = \tfrac{2}{3}(1-F_{\mathrm{proc}})$ for $d=2$.
The experimental Choi matrix is estimated by maximum-likelihood estimation (MLE) constrained to completely-positive trace-preserving (CPTP) maps.
The dominant eigenvalue of $\mathcal{C}$ identifies the closest unitary channel, which captures the coherent part of the error; the remaining eigenvalues encode incoherent contributions such as dephasing and SPAM.

\section{Rb analysis}
\label{sec:rb_analysis}
Single qubit randomized benchmarking provides an independent estimate of the fidelity per Clifford gate that is, by construction, insensitive to state-preparation and measurement (SPAM) errors when the full depolarising model $f(m) = A\,\alpha^m + b$ is fitted. In practice, the asymptote $b$ and amplitude $A$ absorb residual SPAM, so the naive fit biases the estimate of the true gate error. To isolate the gate contribution, the SPAM parameters are determined \emph{offline} by fitting the digital-twin model to the full experimental RB curve (Table \ref{tab:spam_ds1_ds2}); these calibrated SPAM offsets are then held fixed while the decay rate $\alpha$ is re-fitted, yielding the SPAM-corrected error per Clifford $\mathrm{EPC} = \tfrac{1}{2}(1-\alpha)$.

\begin{figure}
    \centering
    \includegraphics[width=0.6\linewidth]{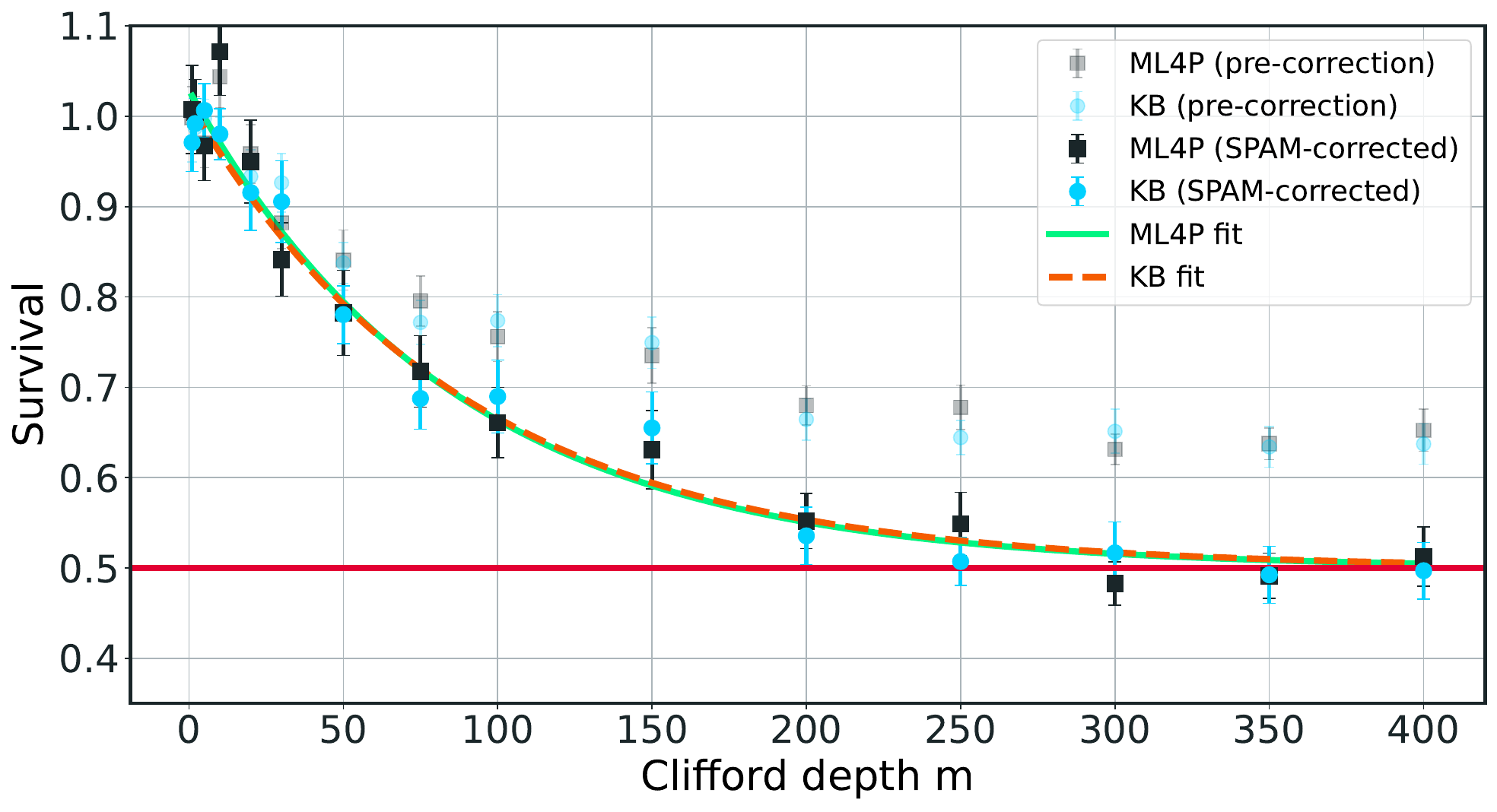}
    \caption{Single-qubit RB survival-probability curves for KB and ML4P learned parameteres. Pre-correction  (raw) points are shown as well as the post-correction ones. SPAM parameters are obtained by offline model learning of selected RB curve points using the digital twin, isolating the intrinsic gate depolarisation rate $\alpha$ and yielding to the error per Clifford $\mathrm{EPC}=\tfrac{1}{2}(1-\alpha)$.}
    \label{fig:rb-corrected_and_NOT}
\end{figure}

In Figure~\ref{fig:rb-corrected_and_NOT} shows the RB survival-probability curve before and after the SPAM correction is applied. The pre-correction data (raw) sits below the post-correction curve at long sequence lengths because the fitted asymptote $b$ absorbs readout errors; once SPAM is removed the corrected decay rate reflects the intrinsic gate fidelity.

\begin{table}[t]
\centering
\begin{tabular}{lcc}
\toprule
 & ML4P & KB \\
\midrule
\multicolumn{3}{l}{\textit{Learned SPAM / dephasing parameters}} \\
$T_2^{*}$ ($\mu$s)            & $35.7 \pm 7.6$              & $35.1 \pm 5.6$ \\
$p_{0\to1}$                    & $0.000 \pm 0.017$ & $0.000 \pm 0.013$ \\
$p_{1\to0}$                    & $0.287 \pm 0.040$           & $0.279 \pm 0.030$ \\
\midrule
\multicolumn{3}{l}{\textit{Per-Clifford infidelity}} \\
$r_{\rm raw}$ (naive fit)          & $(5.30 \pm 0.83)\times10^{-3}$ & $(5.11 \pm 0.76)\times10^{-3}$ \\
$r_{\rm gate}$ (SPAM-corrected)    & $(5.82 \pm 0.62)\times10^{-3}$ & $(5.62 \pm 0.54)\times10^{-3}$ \\
$F_{\rm gate}$ (\%)                & $99.4178$ & $99.4377$ \\
\bottomrule
\end{tabular}
\caption{Learned SPAM/dephasing parameters and RB fit results for ML4P and KB. Parameter uncertainties are 1$\sigma$ calculated via GN estimation.
$\chi^2_\nu$ is the reduced chi-square of the corresponding exponential fit ($\nu$ = degrees of freedom).}
\label{tab:spam_ds1_ds2}
\end{table}

\bibliographystyle{unsrt}
\bibliography{references}

\end{document}